\documentclass[twocolumn]{aa}      
\usepackage{graphicx}      

\begin{document}      
   \title{Mass motions and chromospheres of RGB stars in the globular 
   cluster NGC 2808     
   \thanks{Based on observations collected at the European Southern      
   Observatory, Chile, during FLAMES Science Verification}}      
      
      
   \author{C. Cacciari\inst{1},     
   	 A. Bragaglia\inst{1},      
	 E. Rossetti\inst{1},      
	 F. Fusi Pecci\inst{1},      
	 G. Mulas\inst{2},     
	 E. Carretta\inst{3},    
	 R.G. Gratton\inst{3},	 	       
	 Y. Momany\inst{4}      
	 and     
	 L. Pasquini\inst{5}     
	 }      
      
   \offprints{C. Cacciari}      
      
   \institute{     
           INAF--Osservatorio Astronomico di Bologna, via Ranzani 1,      
            I--40126 Bologna, Italy \\      
           \email{cacciari@bo.astro.it, angela@bo.astro.it, flavio@bo.astro.it}	         
	     \and      
	   INAF--Osservatorio Astronomico di Cagliari,  Loc. Poggio dei Pini,      
	   Strada 54, I--09012 Capoterra (CA), Italy \\     
	    \email{gmulas@ca.astro.it}     
  	      \and     
	   INAF--Osservatorio Astronomico di Padova, vicolo Osservatorio 5,      
            I--35122 Padova, Italy \\      
           \email{carretta@pd.astro.it, gratton@pd.astro.it}     
              \and     
           Universit\`a di Padova, Dip. di Astronomia, vicolo Osservatorio 2,     
	   I--35122 Padova, Italy \\      
           \email{momany@pd.astro.it}     
              \and     
	   European Southern Observatory,      
	   Karl-Schwarzschild-Str. 2, D-85749 Garching b. M\"unchen,      
	   Germany \\     
	   \email{lpasquin@eso.org}	     	        
	   }       
     
   \date{}      
      
   \abstract{We present the results of the first observations, taken with      
   FLAMES during Science Verification, of red giant branch (RGB) stars      
   in the globular cluster NGC 2808. 
   A total of 137 stars was observed, of which 20 at high      
   resolution (R=47,000) with UVES and the others at lower resolution      
   (R=19,000-29,000) with GIRAFFE in MEDUSA mode, 
   monitoring $\sim$ 3 mag down from the RGB 
   tip. Spectra were taken of the H$\alpha$,  Na {\sc i} D and Ca {\sc ii} H 
   and K lines. 
   This is by far the largest and most complete collection of such data in 
   globular cluster giants, both for the number of stars observed within one 
   cluster, and for monitoring all the most important optical diagnostics of 
   chromospheric activity/mass motions.
   Evidence of mass motions in the atmospheres was searched from asymmetry in 
   the profiles and coreshifts of the H$\alpha$,  Na {\sc i} D and Ca {\sc ii} 
   K lines, as well as from H$\alpha$ emission wings.     
   We have set the detection thresholds for the onset of H$\alpha$ emission, 
   negative Na $D_2$ coreshifts and negative $K_3$ coreshifts at 
   $\log L/L_{\odot}\sim$2.5, 2.9 and 2.8, respectively. These limits are 
   significantly fainter than the results found by nearly all previous studies. 
   Also the fraction of stars where these features have been detected has 
   been increased significantly with respect to the previous studies. 
   Our observations confirm the widespread presence of 
   chromospheres among globular cluster giants, as it was found 
   among Population I red giants. Some of the above 
   diagnostics suggest clearly the presence of outward mass motions in the 
   atmosphere of several stars.

   }      
      
 \authorrunning{Cacciari et al.}      
 \titlerunning{Chromospheres/mass motions in NGC 2808 red giant stars}      
 \maketitle      
%
      
\section{Introduction}

One of the most solid requirements of the stellar evolution theory      
is that some amount of mass loss (a few tenths of a solar mass) must occur      
during the evolutionary phases preceding the Horizontal Branch (HB) phase,      
in order to account for the observed HB morphologies in globular clusters (GC)      
(Castellani \& Renzini 1968; Iben \& Rood 1970; Rood 1973;      
Fusi Pecci \& Renzini 1975, 1976; Renzini 1977).      
Also the pulsational properties of the RR Lyrae      
variables and the absence of asymptotic giant branch (AGB) stars brighter      
than the red giant branch (RGB) tip require that some mass has been      
lost during previous evolutionary phases      
(Christy 1966; Fusi Pecci et al. 1993; D'Cruz et al. 1996).      
Theoretical estimates of mass loss rates at the tip of the RGB are a few      
times $10^{-8}$ M$_{\odot}$ yr$^{-1}$ (Fusi Pecci \& Renzini 1975, 1976;      
Renzini 1977). This would produce a few tens of solar masses of      
intracluster matter that should accumulate in the central regions of the      
clusters, in absence of sweeping mechanisms between Galactic plane      
crossings.      
     
However, efforts to obtain      
direct evidence of intracluster matter, or of mass loss      
from individual RGB stars, have been only marginally successful.      
Diffuse gas in GC's  was detected      
only as an upper limit and well below 1 $M_{\odot}$ (Roberts 1988;      
Smith et al. 1990; Faulkner \& Smith 1991; Freire et al. 2001),      
whereas the most recent search      
of mass loss evidence from individual RGB (or AGB) stars via ISOCAM      
IR-excess has indeed shown that dusty circumstellar envelopes are present      
in $\sim$ 15\% of the  giants in the $\sim$ 0.7 mag brightest interval      
(M$_{\rm bol} \le -2.5$)      
(Origlia et al. 2002, and references therein).      
     
Spectroscopic surveys of a few hundred GC red giants      
(Cohen 1976, 1978, 1979, 1980, 1981; Mallia \& Pagel 1978;      
Peterson 1981, 1982; Cacciari \& Freeman 1983; Gratton et al. 1984)      
did reveal H$\alpha$ emission wings in a good fraction of stars brighter      
than log L/L$_{\odot}\sim$ 2.7, i.e. along the upper 1.25 mag interval      
of the RGB. This was initially interpreted as evidence of an extended    
atmosphere, i.e. of mass loss. However, Dupree et al. (1984) demonstrated    
that this emission   {\it per se} is not a unambiguous mass loss indicator,    
as it     could arise naturally in a static stellar chromosphere, or it    
could be   influenced by hydrodynamic processes due to pulsation    
(Dupree et al. 1994).     
     
Profile asymmetry and coreshifts of chromospheric lines can reveal mass      
motions, and in particular the presence of a stellar wind and circumstellar     
material. Red giants in globular clusters were found to exhibit low velocity      
shifts in the cores of the H$\alpha$ or Na {\sc i} D lines (cf. Peterson 1981;     
Bates et al. 1990, 1993); similarly, metal-poor field giants, which might be      
taken as the field counterparts of globular cluster giants, also indicate      
slow outflow from the asymmetries and line shifts in the  H$\alpha$, Ca {\sc ii} and     
Mg {\sc ii} lines (Smith et al. 1992; Dupree \& Smith 1995).      
      
More recently, Lyons et al. (1996) discussed the Na {\sc i} D and H$\alpha$ 
stellar     
profiles for a sample of 63 RGB stars in 5 GC (M4, M13, M22, M55 and      
$\omega$ Cen), and found evidence of significant Na {\sc i} D core shifts in     
$\sim$ 50\% of the stars brighter than $\log L/L_{\odot}\sim$2.9, whereas      
significant H$\alpha$ core shifts were detected in      
$\sim$ 50\% of the stars brighter than $\log L/L_{\odot}\sim$2.5. These      
coreshifts are all $\le$ 10 kms$^{-1}$, i.e. much smaller than the escape      
velocity from the stellar photosphere ($\sim$ 50-60 kms$^{-1}$).      
     
Two RGB stars in NGC 6752 were studied  by Dupree et al. (1994),      
by a detailed analysis of the Mg {\sc ii}, Ca {\sc ii} K and H$\alpha$ line profiles.     
These stars are at the RGB tip, and the Ca {\sc ii} K and H$\alpha$ core shifts     
again revealed slow ($\le$ 10 kms$^{-1}$) outflow motions. The asymmetries in      
the Mg {\sc ii} lines, however, indicated under certain assumptions a stellar      
wind with a terminal      
velocity of $\sim$ 150 kms$^{-1}$, exceeding both the stellar photospheric      
escape velocity ($\sim$ 55 kms$^{-1}$) and the escape velocity from the      
cluster core ($\sim$ 23 kms$^{-1}$). The mass loss rate estimated by      
Dupree et al. (1994) from the Mg {\sc ii} results      
($\sim 10^{-9}$ M$_{\odot}$ yr$^{-1}$) would lead to a total mass loss of      
$\sim$ 0.2$ M_{\odot}$ over the star lifetime on the RGB      
($\sim 2 \times 10^{8}$ yr), in very good agreement with the expectations      
of the stellar evolution theory. This result alone is not sufficient to      
meet the requirements of the stellar evolution that {\em all} stars suffer      
{\em some degree} of mass loss during the phases preceding the HB. However,      
it shows that the mass loss phenomenon along the RGB does indeed occur,      
even if perhaps only occasionally and detected among the brightest stars,      
and it may be revealed using visual indicators, although less effectively and      
accurately than using chromospheric lines in the UV such as the Mg {\sc ii}, or in      
the near-IR such as the He {\sc i} at 10830 \AA ~(Dupree et al. 1992).      
      
The advent of the multi-fibre spectrograph FLAMES on VLT2-Kueyen      
(Pasquini et al. 2002),      
with a multiplex capability of 8 with UVES (R=45000) and $\sim$ 130 with      
GIRAFFE in MEDUSA mode (R$\sim$15000--30000), allows a      
much more efficient monitoring of visual diagnostics of mass outflow along      
the RGB over a large magnitude range down from the RGB tip, especially in      
terms of sample size.        
     
We have selected the globular cluster NGC 2808 for this monitoring, as      
it was the best candidate available for January-February observations, when      
Science Verification (SV) observations were scheduled.             
       
\begin{table*}      
\begin{center}      
\label{t:abu1}      
\caption[]{List of RGB stars observed with FLAMES during Science Verification      
 on Jan 24 -- Feb 02, 2003. Identification, coordinates and visual photometry      
 are taken from Piotto et al. (2003), IR photometry is taken from the      
 2MASS Catalog. }      
\begin{tabular}{rrrcccccl}      
\hline\hline      
\\      
ID n. & RA(2000) & DEC(2000) & B & V & J & H & K & Notes  \\      
\\      
\hline      
\\      
 7183 &9 12 02.8710 &-64 49 34.069&16.168 &14.854 &	  &	  &	   &2 \\     
 7536 &9 12 31.7065 &-64 49 22.268&15.812 &14.372 &11.802 &11.069 &10.829  &1,2 \\     
 7558 &9 12 20.1287 &-64 49 21.891&16.576 &15.389 &13.160 &12.518 &12.407  &1 \\     
 7788 &9 11 57.1979 &-64 49 14.551&16.168 &14.870 &12.472 &11.777 &11.623  &1,2 \\     
 8603 &9 12 14.0510 &-64 48 42.915&15.847 &14.432 &11.902 &11.162 &10.957  &1 \\     
 8679 &9 11 44.5585 &-64 48 39.890&16.164 &14.961 &12.753 &12.156 &12.022  &1,2,F \\     
 8739 &9 11 51.2018 &-64 48 37.535&15.761 &14.288 &11.620 &10.813 &10.693  &1,2 \\     
 8826 &9 12 09.8073 &-64 48 33.243&16.275 &15.033 &12.746 &12.057 &11.911  &1,2 \\     
 9230 &9 12 13.3528 &-64 48 11.617&15.591 &14.028 &11.252 &10.449 &10.303  &1 \\     
 9724 &9 12 00.9163 & 64 47 43.204&15.869 &14.508 &11.994 &11.282 &11.090  &1,2 \\     
 9992 &9 12 46.8254 &-64 47 23.732&15.402 &13.702 &10.729  &9.886 & 9.659  &2 \\     
10012 &9 11 58.5661 &-64 47 23.138&15.682 &14.225 &11.613 &10.836 &10.679  &1,2 \\     
10105 &9 12 21.1477 &-64 47 13.906&15.976 &14.669 &12.279 &11.564 &11.417  &1,2 \\     
10201 &9 12 45.7573 &-64 47 05.388&16.864 &15.714 &13.508 &12.868 &12.759  &1,3 \\     
10265 &9 11 56.6812 &-64 47 00.535&15.728 &14.309 &11.772 &11.026 &10.823  &1,2 \\     
10341 &9 12 25.4161 &-64 46 52.409&16.009 &14.659 &12.205 &11.526 &11.335  &1,2 \\     
10571 &9 12 41.1223 &-64 46 25.848&15.806 &14.376 &11.771 &10.988 &10.844  &1,2 \\     
10681 &9 12 05.7235 &-64 46 13.801&15.261 &13.545 &10.635 & 9.855 & 9.586  &2 \\     
13575 &9 11 31.6300 &-64 49 02.281&16.615 &15.386 &13.090 &12.417 &12.270  &1 \\     
13983 &9 11 39.7915 &-64 48 25.463&17.027 &15.940 &13.832 &13.246 &13.092  &1,3 \\     
30523 &9 11 22.4875 &-64 55 23.351&16.658 &15.489 &13.320 &12.697 &12.550  &1 \\     
30763 &9 11 31.6195 &-64 54 57.989&15.927 &14.606 &12.287 &11.577 &11.465  &2 \\     
30900 &9 11 36.3238 &-64 54 47.078&16.319 &15.084 &12.890 &12.203 &12.075  &1,2 \\     
30927 &9 11 41.3502 &-64 54 45.153&15.215 &13.516 &10.814 & 9.985 & 9.802  &2 \\     
31851 &9 11 14.3696 &-64 53 36.046&16.420 &15.221 &13.064 &12.411 &12.280  &1 \\     
32398 &9 11 39.4612 &-64 53 01.270&16.004 &14.687 &12.321 &11.590 &11.483  &2 \\     
32469 &9 11 16.3279 &-64 52 55.955&15.485 &14.040 &11.531 &10.777 &10.689  &1 \\     
32685 &9 11 27.0742 &-64 52 44.375&16.794 &15.656 &13.548 &12.956 &12.772  &1,3 \\     
32862 &9 11 42.0399 &-64 52 34.140&16.379 &15.166 &12.871 &12.171 &11.999  &1,2 \\     
32909 &9 11 39.6760 &-64 52 31.924&16.272 &15.022 &12.710 &12.046 &11.910  &1\\     
32924 &9 11 22.2965 &-64 52 31.081&16.138 &14.853 &12.515 &11.785 &11.701  &1 \\     
33452 &9 11 40.7982 &-64 52 01.561&16.405 &15.170 &12.970 &12.271 &12.167  &2 \\     
33918 &9 11 01.6928 &-64 51 36.087&15.776 &14.331 &11.756 &11.021 &10.887  &1,2 \\     
34008 &9 11 27.5242 &-64 51 31.290&16.346 &15.119 &12.894 &12.240 &12.115  &2 \\     
34013 &9 11 23.8497 &-64 51 30.965&17.499 &16.444 &14.503 &13.897 &13.843  &3 \\          
35265 &9 11 22.7197 &-64 50 12.117&16.584 &15.321 &12.965 &12.287 &12.180  &1 \\       
37496 &9 12 13.8815 &-64 57 09.411&16.126 &14.871 &12.586 &11.891 &11.730  &1,2 \\     
37505 &9 12 26.4217 &-64 57 08.391&16.058 &14.783 &12.471 &11.739 &11.591  &2 \\     
37781 &9 12 13.0342 &-64 56 41.142&16.148 &15.037 &13.001 &12.396 &12.307  &1,F \\     
37872 &9 12 23.0712 &-64 56 34.303&15.334 &13.650 &10.767  &9.904 & 9.711  &2,3 \\     
37998 &9 12 12.5554 &-64 56 24.603&15.638 &14.233 &11.706 &10.960 &10.796  &1,2 \\     
38228 &9 12 30.9696 &-64 56 08.521&16.211 &14.981 &12.700 &12.035 &11.887  &1 \\     
38244 &9 12 08.9333 &-64 56 07.955&16.378 &15.193 &12.996 &12.307 &12.177  &1 \\     
38559 &9 12 56.1453 &-64 55 48.356&15.541 &14.133 &11.674 &10.915 &10.754  &1,2,F \\     
38660 &9 12 39.8673 &-64 55 43.083&15.772 &14.452 &11.995 &11.407 & 9.519  &2 \\     
38967 &9 12 08.8682 &-64 55 27.764&16.415 &15.252 &13.060 &12.383 &12.297  &1 \\     
39060 &9 12 14.5222 &-64 55 24.085&15.793 &14.417 &11.939 &11.193 &11.036  &1 \\     
39577 &9 12 10.9820 &-64 55 03.959&15.820 &14.509 &12.085 &11.357 &11.189  &1,2,F \\     
40196 &9 12 05.9806 &-64 54 43.265&15.894 &14.570 &12.142 &11.440 &11.264  &2 \\     
40983 &9 11 48.6726 &-64 54 20.737&15.723 &14.360 &11.906 &11.153 &10.988  &1,2 \\     
41008 &9 12 17.6744 &-64 54 19.865&16.339 &15.189 &12.973 &12.309 &12.200  &1 \\     
41828 &9 12 08.5555 &-64 54 00.670&15.862 &14.521 &12.129 &11.316 &11.210  &1 \\     
41969 &9 11 50.8999 &-64 53 57.615&15.787 &14.446 &11.972 &11.254 &11.092  &1,2 \\     
42073 &9 12 16.2176 &-64 53 55.186&15.702 &14.291 &11.733 &11.031 &10.783  &1 \\     
 \\     
\hline      
\end{tabular}      
\end{center}      
\label{tab1}     
\end{table*}      
      
\addtocounter{table}{-1}     
     
\begin{table*}      
\begin{center}      
\label{t:abu1}      
\caption[]{(Continuation)}       
\begin{tabular}{rrrcccccl}      
\hline\hline      
\\      
ID n. &RA(2000) &DEC(2000) &B &V &J &H &K &Notes  \\      
     
\\      
\hline      
\\       
42165 &9 12 33.7942 &-64 53 52.922&17.430 &16.387 &14.369 &13.816 &13.613 & 1 \\     
42789 &9 11 48.9406 &-64 53 41.364&16.476 &15.367 &13.246 &12.610 &12.507 & 1 \\     
42886 &9 12 59.1095 &-64 53 38.608&17.019 &15.921 &13.780 &13.157 &13.040 & 1,3 \\     
42996 &9 12 25.0560 &-64 53 37.226&16.552 &15.426 &13.294 &12.656 &12.523 & 1 \\     
43041 &9 12 40.0467 &-64 53 35.930&15.349 &13.723 &10.869 &10.063 & 9.822 & 1,2 \\     
43217 &9 12 32.6714 &-64 53 32.870&17.471 &16.440 &14.454 &13.805 &13.675 & 3 \\     
43247 &9 12 10.1184 &-64 53 32.654&16.022 &14.786 &12.486 &11.732 &11.543 & 1,2 \\     
43281 &9 12 21.2449 &-64 53 32.058&16.270 &15.047 &	  &	 &	  & 2 \\     
43333 &9 12 55.0733 &-64 53 30.392&16.570 &15.473 &13.408 &12.776 &12.698 & 1,F \\     
43561 &9 12 14.4769 &-64 53 26.725&15.162 &13.323 &10.177 & 9.295 & 9.073 & 2 \\     
43794 &9 12 26.3414 &-64 53 22.607&15.861 &14.554 &12.157 &11.468 &11.271 & 1,2 \\     
43822 &9 11 58.4947 &-64 53 22.341&15.931 &14.600 &12.176 &11.422 &11.291 & 1,2 \\     
44573 &9 12 06.1145 &-64 53 08.983&16.285 &15.089 &12.709 &12.022 &11.912 & 2 \\     
44665 &9 12 15.2786 &-64 53 07.367&15.649 &14.245 &11.718 &10.980 &10.799 & 1,2 \\     
44716 &9 12 17.8493 &-64 53 06.449&16.123 &14.899 &12.605 &11.924 &11.748 & 2 \\     
44984 &9 12 45.8780 &-64 53 01.411&15.855 &14.535 &12.088 &11.340 &11.219 & 1,2 \\     
45162 &9 11 59.2356 &-64 52 59.262&15.172 &13.263 &10.019 & 9.074 & 8.853 & 2 \\     
45443 &9 12 21.3723 &-64 52 54.137&15.885 &14.574 &12.128 &11.385 &11.283 & 1,2 \\     
45840 &9 12 11.1152 &-64 52 47.477&15.552 &14.117 &11.516 &10.703 &10.536 & 1,2 \\          
46041 &9 12 02.9671 &-64 52 44.370&16.166 &14.967 &12.683 &12.008 &11.850 & 1,2 \\     
46099 &9 12 33.5788 &-64 52 43.088&15.375 &13.741 &10.915 &10.044 & 9.835 & 2,3 \\     
46367 &9 12 00.4581 &-64 52 38.810&16.054 &14.832 &12.526 &11.884 &11.485 & 1 \\     
46422 &9 11 56.0932 &-64 52 37.896&15.174 &13.376 &	  &	  &	  &3\\     
46580 &9 11 56.1904 &-64 52 35.387&15.292 &13.690 &	  &	  &	  &2,3 \\     
46663 &9 12 17.0998 &-64 52 33.946&15.860 &14.551 &12.146 &11.456 &11.323 & 1 \\     
46726 &9 11 50.5738 &-64 52 33.000&15.740 &14.375 &11.910 &11.160 &10.998 & 2 \\     
46868 &9 12 09.7828 &-64 52 30.452&16.003 &14.752 &	  &	  &	  &1\\     
46924 &9 12 04.6008 &-64 52 29.587&16.096 &14.826 &	  &	  &	  &2\\     
47031 &9 12 00.9644 &-64 52 27.851&15.605 &14.181 &11.593 &10.894 &10.355 & 2 \\     
47145 &9 11 58.1949 &-64 52 25.936&15.249 &13.647 &10.814 & 9.995 & 9.777 & 1,2 \\     
47421 &9 11 56.4377 &-64 52 21.270&15.571 &14.175 &11.674 &10.834 &10.696 & 1,2 \\     
47452 &9 12 20.2257 &-64 52 20.564&15.857 &13.777 &10.268 & 9.499 & 9.157 & 2 \\     
47606 &9 12 06.6571 &-64 52 18.225&15.257 &13.436 &10.227 & 9.349 & 9.103 & 2,3 \\     
48011 &9 11 48.0108 &-64 52 12.236&15.718 &14.397 &11.965 &11.233 &11.040 & 1,2 \\     
48060 &9 11 55.1995 &-64 52 11.458&15.256 &13.556 &10.609 & 9.741 & 9.523 & 1,2 \\     
48128 &9 11 52.1753 &-64 52 10.603&15.825 &14.484 &11.997 &11.292 &11.118 & 1,2 \\     
48424 &9 12 06.6195 &-64 52 05.871&15.772 &14.452 &11.995 &11.407 & 9.519 & 2 \\     
48609 &9 12 16.6443 &-64 52 02.923&15.285 &13.415 &10.229 & 9.353 & 9.101 & 3 \\     
48889 &9 12 08.5086 &-64 51 58.467&15.139 &13.341 &10.340 & 9.431 & 9.253 & 2,3 \\     
49509 &9 11 56.2108 &-64 51 49.116&15.429 &13.921 &11.237 &10.379 &10.250 & 1 \\     
49680 &9 12 10.0583 &-64 51 46.474&15.256 &13.567 &10.573 & 9.706 & 9.497 & 1 \\     
49743 &9 12 36.7974 &-64 51 45.097&16.353 &15.168 &12.925 &12.266 &12.159 & 1,2 \\     
49753 &9 12 06.8961 &-64 51 45.474&15.669 &14.350 &11.804 &11.018 &10.886 & 2 \\     
49942 &9 12 12.0864 &-64 51 42.617&15.400 &13.736 &10.848 &10.027 & 9.812 & 1 \\     
50119 &9 11 42.9184 &-64 51 39.752&15.425 &13.886 &11.210 &10.382 &10.218 & 2,3 \\     
50371 &9 12 07.9720 &-64 51 35.858&15.464 &13.840 &10.999 &10.144 & 9.976 & 1 \\     
50561 &9 12 23.1839 &-64 51 32.695&16.120 &14.834 &12.428 &11.725 &11.548 & 2 \\     
50681 &9 12 18.4772 &-64 51 30.760&15.303 &13.660 &10.874 &10.100 & 9.905 & 2 \\     
50761 &9 11 57.0729 &-64 51 29.676&15.394 &13.390 & 9.947 & 9.096 & 8.827 & 2,3 \\     
50861 &9 12 11.0865 &-64 51 27.889&15.201 &13.556 &10.727 & 9.851 & 9.662 & 1,2 \\     
50866 &9 11 54.2702 &-64 51 27.928&15.941 &14.634 &12.219 &11.524 &11.353 & 1 \\     
50910 &9 11 42.6346 &-64 51 27.221&16.234 &14.984 &12.697 &12.033 &11.877 & 1 \\     
51416 &9 12 29.2944 &-64 51 18.935&16.188 &14.945 &12.619 &11.911 &11.803 & 1,2 \\     
51454 &9 12 02.2715 &-64 51 18.513&15.233 &13.446 &10.336 & 9.492 & 9.244 & 2,3 \\     
51499 &9 12 07.3496 &-64 51 17.800&15.157 &13.435 &10.447 & 9.567 & 9.382 & 2,3 \\     
51515 &9 11 59.3832 &-64 51 17.525&15.812 &14.213 &11.472 &10.665 &10.525 & 1,2 \\     
51646 &9 11 50.3432 &-64 51 15.395&16.550 &15.460 &13.425 &12.791 &12.668 & 1 \\     
51871 &9 12 05.4500 &-64 51 11.901&15.309 &13.542 &10.040 & 9.718 & 9.514 &1 \\     
 \\     
\hline      
\end{tabular}      
\end{center}      
\end{table*}      
     
\addtocounter{table}{-1}      
     
\begin{table*}      
\begin{center}      
\label{t:abu1}      
\caption[]{(Continuation)}       
\begin{tabular}{rrrcccccl}      
\hline\hline      
\\      
ID n. &RA(2000) &DEC(2000) &B &V &J &H &K &Notes  \\      
\\      
\hline      
\\       
51930 &9 11 49.9340 &-64 51 10.872&15.531 &14.068 &11.422 &10.655  &10.478 &2 \\     
51963 &9 12 27.6355 &-64 51 10.138&15.652 &14.199 &11.577 &10.821  &10.652 &1,2 \\     
51983 &9 12 02.5038 &-64 51 10.068&15.328 &13.474 &10.294 & 9.391  & 9.182 &3 \\     
52006 &9 12 18.1017 &-64 51 09.524&15.787 &14.446 &11.972 &11.254  &11.092 &1,2 \\     
52048 &9 12 05.7704 &-64 51 80.996&15.645 &14.259 &10.387 &11.012  &10.797 &2 \\     
52647 &9 12 12.2817 &-64 50 59.590&15.925 &14.654 &12.293 &11.592  &11.432 &2 \\     
53284 &9 12 01.4868 &-64 50 49.350&15.501 &13.847 &10.902 & 9.254  & 9.953 &2 \\     
53390 &9 11 52.9455 &-64 50 47.860&15.958 &14.669 &12.271 &11.573  &11.379 &3 \\     
53579 &9 11 48.7436 &-64 50 44.801&15.830 &14.461 &11.877 &11.178  &10.992 &2 \\     
54264 &9 12 21.5127 &-64 50 33.737&15.285 &13.415 &10.229 & 9.353  & 9.101 &1 \\     
54284 &9 11 54.7689 &-64 50 33.662&15.789 &14.407 &11.877 &11.148  &10.985 &1 \\     
54308 &9 12 01.6780 &-64 50 33.286&15.568 &13.659 &10.182 & 9.264  & 9.024 &2 \\     
54733 &9 12 07.8037 &-64 50 26.397&15.602 &14.128 &11.523 &10.761  &10.571 &1,2 \\     
54756 &9 11 46.3335 &-64 50 26.044&16.086 &14.784 &12.416 &11.679  &11.528 &2 \\     
54789 &9 12 00.4239 &-64 50 25.567&15.626 &14.247 &11.756 &11.019  &10.863 &1,2 \\     
55031 &9 11 58.1972 &-64 50 21.703&15.283 &13.600 &10.680 & 9.813  & 9.620 &1 \\     
55354 &9 12 10.3337 &-64 50 16.108&16.024 &14.749 &12.406 &11.655  &11.531 &2 \\     
55437 &9 12 11.4613 &-64 50 14.694&15.982 &14.692 &12.302 &11.586  &11.420 &1 \\     
55609 &9 12 29.9315 &-64 50 11.384&16.078 &14.751 &12.301 &11.575  &11.467 &1,2 \\     
55627 &9 12 05.4152 &-64 50 11.404&15.899 &14.585 &12.188 &11.524  &11.338 &1 \\     
56032 &9 11 45.5811 &-64 50 04.209&15.579 &13.976 &11.112 &10.292  &10.095 &3 \\     
56136 &9 12 30.9806 &-64 50 02.010&17.408 &16.383 &14.434 &13.879  &13.608 &1 \\     
56536 &9 12 04.3411 &-64 49 55.639&15.301 &13.626 &10.785 & 9.896  & 9.688 &2 \\     
56710 &9 12 30.1594 &-64 49 52.049&16.582 &15.430 &13.232 &12.553  &12.396 &1 \\     
56924 &9 12 15.1423 &-64 49 48.859&15.629 &14.207 &11.587 &10.869  &10.676 &2 \\     
\hline      
\end{tabular}      
\end{center}      
Notes: 1=MEDUSA HR12 + HR14; 2=MEDUSA HR02; 3=UVES; F= field star.       
\end{table*}      
     
\begin{figure}      
\begin{center}     
\includegraphics[width=9cm]{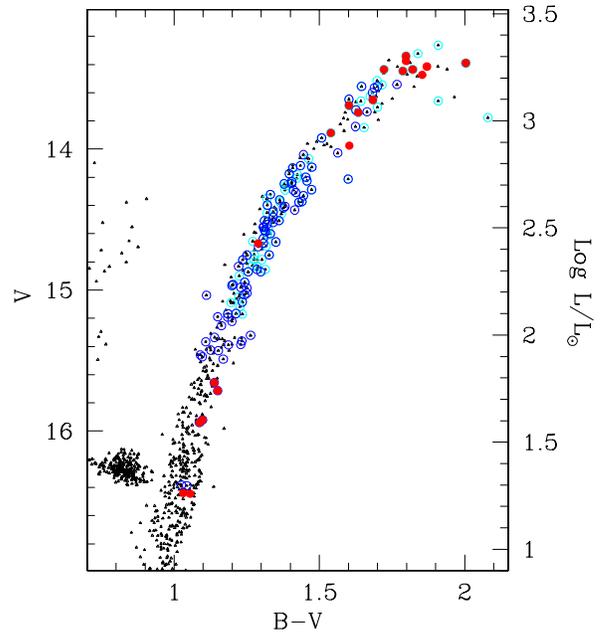}       
\caption{Colour-Magnitude diagram of NGC2808 (Piotto et al. 2003)     
showing the 137 RGB stars observed      
with FLAMES. Filled circles indicate the stars observed with UVES,      
open circles indicate the stars observed with GIRAFFE/MEDUSA.}      
\label{figcmd}      
\end{center}     
\end{figure}      
     
\begin{figure*}     
\begin{center}      
\includegraphics[width=18cm,height=9cm]{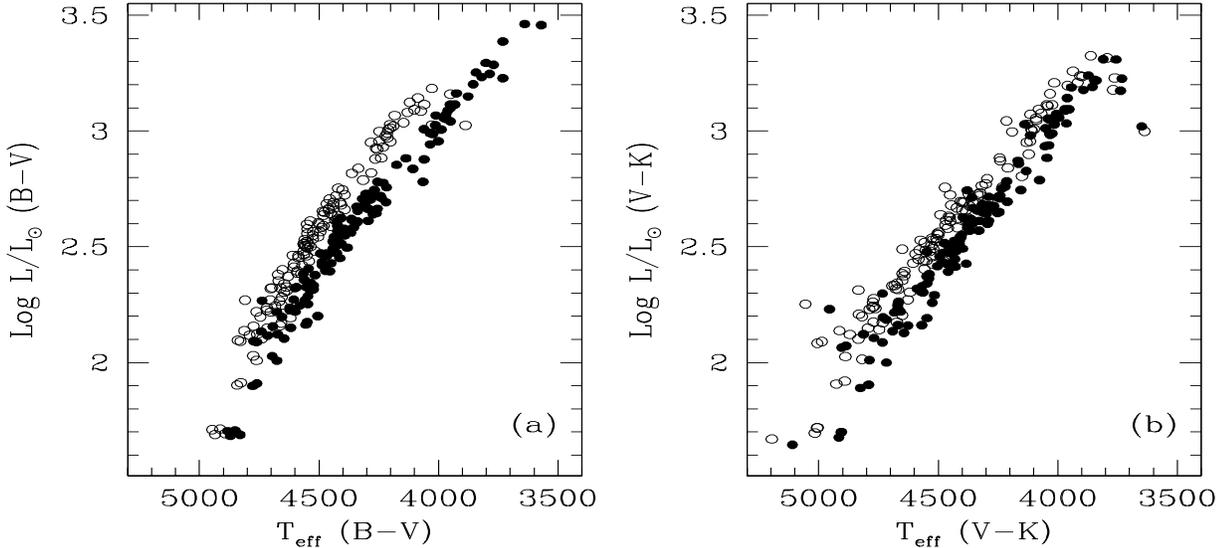}       
\caption{Comparison of the temperatures and luminosities     
obtained using the Montegriffo et al.      
(1998, open circles) and Alonso et al. (1999,2001 filled dots) calibrations,      
for the $(B-V)$ colours (panel a) and $(V-K)$ colours (panel b). }      
\label{te_cals}      
\end{center}     
\end{figure*}

\begin{figure}     
\begin{center}      
\includegraphics[width=9cm,height=9cm]{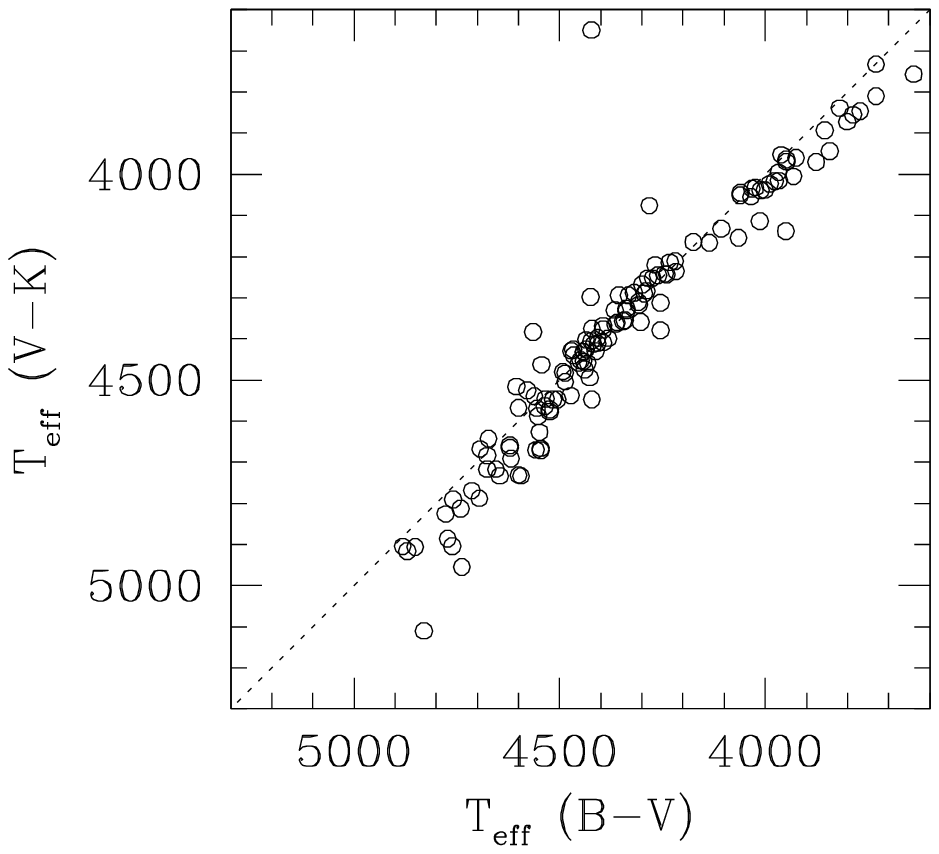}       
\caption{Comparison of the temperatures obtained from the $(B-V)$ and     
$(V-K)$ colours, using Alonso et al. (1999,2001) calibration.   }      
\label{tes}      
\end{center}     
\end{figure}      
 
\section{Observations and data reduction}      
     
The target selection and observation were made possible by the accurate UBV      
photometry and astrometry provided by Piotto et al. (2003).   All stars have     
been  checked to be free from companions closer than 2.4 arcsec and brighter     
than  V+1.5, where V is the target magnitude. NGC 2808 is quite     
concentrated, and all target stars lie within a 7\arcmin     
~radius from the cluster centre, and this has put to a difficult test the     
capabilities of the FLAMES fiber positioner (FLAMES has a corrected field     
of view of 25\arcmin ~diameter).     
     
We show in Fig. \ref{figcmd} the Color-Magnitude diagram of NGC 2808: 
the tip of the RGB is at $V$ = 13.2, corresponding to       
M$_V \sim $ --2.3, T$_{\rm eff}\sim$3800 K, log L/L$_{\odot}\sim$ 3.3, and      
M$_{bol} \sim $ --3.5.      
We have selected 13 RGB stars within the uppermost 0.8 mag      
interval and 7 more sampling 2.5 fainter magnitudes, to be observed with      
UVES (R=47000, 8 single fibres of 1 arcsec entrance aperture, 
grating centred at 580nm and covering about 200nm).      
This wavelength range includes the  Na {\sc i} D and H$\alpha$ lines. 
The exposure times were 1800s for the stars with $V < 14$      
(except stars \#50119, 51499 and 56032 that were observed for 3600s each),      
and $2\times3600s$ for the stars with $V > 14$ (except star \#53390      
that was observed for 3600s).      
Simultaneously, GIRAFFE in MEDUSA mode observed 117 more stars 
along  the entire magnitude range with 3 setups,      
namely HR02 (Ca {\sc ii} H+K, R=19600), HR12 (Na {\sc i} D, R=18700) and      
HR14 (H$\alpha$, R=28800). The size of the MEDUSA fibres on the 
sky is 1.2 arcsec.      
The exposure times were $3\times3600s$ with HR2, and 1800s each      
with HR12 and HR14.       
Not all the stars were observed with all the MEDUSA setups due to the     
high crowding conditions and the difficulty of placing the fibers; about     
one fourth were observed only with the HR02 or HR12+HR14 setups, but more than     
50 have complete coperture of the interesting spectral features, either     
entirely with the GIRAFFE setups, or in combination with the UVES spectra.     
Table \ref{tab1} gives in the last column information on the setups used      
for each star.     
We note that this is the first time that such a large sample of 
globular cluster RGB stars have been observed in all the major, deep optical
lines (i.e., H$\alpha$, Na D and Ca {\sc ii} H and K) that are normally used 
to  study the presence of chromospheres and/or mass motions in the atmospheres. 
     
Given the large interval in magnitude, S/N ratios vary a lot.     
In the case of the UVES spectra, S/N measured near 630 nm     
is about 100 for the brighter stars, and about 40 for     
the fainter ones, with 120 and 20 being the extremes.     
In the Ca {\sc ii} region S/N's were quite low (i.e. about 15 at the bottom     
of the K line for the brightest objects), while in the Na {\sc i} D and 
H$\alpha$ setups they varied from about 25 to 150, following the luminosity 
distribution, with peaks around 80-100, and 50-60, respectively. 
        
We collect the photometric information for our target stars in Table 
\ref{tab1},      
including the IR photometry that is now available for all of our targets     
-except 6- in the 2MASS All-Sky Data Release (accessible at  {\em     
www.ipac.caltech.edu/2mass/releases/allsky/} and released on March 25, 2003).      
      
The observations were performed      
during SV between January 24 and February 2, 2003. Most of the data      
were reduced by ESO personnel using the standard pipelines for FLAMES      
(see Pasquini et al. 2002 for a  description), and were made public,       
according to ESO SV policies,  on March 3, 2003.     
These reductions were intended to give a first insight on the data, and     
were not deemed optimal by the SV team: e.g., offsets in wavelegth 
calibrations between fibers could be present, in some cases producing errors 
of up to $\pm$ 10 kms$^{-1}$  in radial velocities (RV's). 
We inspected our spectra and found them quite suitable for our purposes 
in two of the three GIRAFFE setups (HR02 and 12).     
The wavelength calibration for the H$\alpha$ setup HR14 was not quite so
satisfactory,  but no re-reduction of the HR14 spectra was performed 
since the GIRAFFE pipeline was still affected by a number of problems. 
However, for our purpose we mostly need to analyse H$\alpha$ profiles, and 
an accurate wavelength calibration over the entire wavelength range of HR14 
is not essential, as long as it can be trusted over a small wavelength range 
encompassing H$\alpha$ itself and a few close photospheric lines. 
We then defined a pseudo-RV based on 
a few photospheric lines in the immediate vicinity of H$\alpha$, shifted to 
zero RV all the spectra, and retained for further analysis only the 50 \AA  
~centred on  H$\alpha$. No attempt to derive centre/core shifts (see Sect. 4) 
was done on these spectra, but they are perfectly suitable to search 
for emissions. 
Because of these problems with wavelength calibration the  subtraction 
of the sky contribution from the GIRAFFE spectra was not advisable. 
Direct contamination from sky lines could be excluded since the radial velocity
of the cluster is $\sim$ 100 kms$^{-1}$. However, scattered sky light can in
principle affect the line profiles, so we have estimated carefully the counts 
from the sky (using the dedicated fibres) and from the continuum and line core 
at H$\alpha$ for a few stars in the critical magnitude range V=14.0 to 15.0.
For brighter stars the sky contamination becomes irrelevant, and for fainter 
stars no H$\alpha$ emission was detected (see Sect. 4.1.1). 
For homogeneity, we have treated the UVES spectra in the same way. 
We have thus estimated that neglecting the contribution of the sky scattered 
light can introduce an error of $\sim$2\% with UVES at V=14, and of 
$\sim$1-2\% with GIRAFFE at V=14.5-15.0. These are our thresholds for detecting
H$\alpha$ emission, which appears to be much stronger than these values 
(see Sect. 4.1.1), therefore we are confident that neglecting the sky 
contribution in our analysis has not introduced any significant bias in 
our results.  
  
As for the UVES spectra, we decided to reduce them again using the 
available UVES pipeline in order to obtain the best possible accuracy.  
FLAMES/UVES data require a rather careful treatment in order to remove
both instrumental effects common to any high resolution, cross-dispersed 
echelle spectrograph and some specific ones due to its use in multi-object, 
fibre-fed mode. An ad-hoc Data Reduction Software (DRS) was specifically 
developed for this purpose, in the form of a set of procedures which are 
available as a context of the M\"unich Image Data Analysis System (MIDAS). 

This DRS, described in detail elsewhere (Mulas et al. 2002), makes use of 
various flat-field exposures, taken both in fibre and in slit mode, 
to separately 
derive the pixel to pixel response of the detector and the contribution of 
each fibre to the overall distribution of light on the resulting 
two-dimensional science frame. The precise position of the orders and fibres 
is measured on the science frame itself. These pieces of information
are then used together to perform an optimal extraction which, at the same
time, carefully corrects for light contamination between fibres whose dispersed
images are adjacent on the science frame.

Optimal extraction maximises the signal to noise ratio, since it does not 
discard the ``wings'' of the cross-dispersion point spread function (PSF) of 
the fibres, and effectively detects and removes most cosmic ray hits.

Exposures of a reference Th-Ar calibration lamp were extracted in exactly 
the same way as the science frames, to ensure the maximum coherence, and a 
2-dimensional polynomial fit was performed on the detected lines to obtain 
the dispersion function of the spectrograph, separately for each fibre. The 
reference Th-Ar exposures were taken on the same day as the science data to 
be calibrated with them. The resulting wavelength calibration error resulted 
to be well below 0.01~\AA{} throughout the wavelength range. 
Science spectra were rebinned to wavelength space and the echelle orders 
merged, making a weighted average where they overlap. 
Pixel by pixel variances were propagated by the 
DRS throughout the reduction procedure.     
     
Both UVES and GIRAFFE spectra were then analysed using IRAF\footnote{     
IRAF is distributed by the NOAO, which are operated by AURA, under contract     
with NSF}      
and ISA (Gratton 1988) to measure equivalent widths (EW'S) and RV's.          
The RV's were measured from the Doppler shifts of selected photospheric lines,     
mostly of Fe {\sc i}, separately in the GIRAFFE setups (using {\em rvidlines}     
in IRAF on about 20 to 30 lines), and in the UVES spectra (using the ISA package).     
Errors are $\sim $ 0.6 kms$^{-1}$ for UVES and $\sim$ 1.5 kms$^{-1}$     
for GIRAFFE.     
The observed RV's were used to shift all spectra to zero radial velocity,     
thus eliminating any residual possible problem with the zero point shifts     
due to the non optimal calibration, which we found almost non existing anyway 
for the HR02 and 12 setups. Multiple exposures of the same star were 
coadded to enhance the S/N.      
 
In a few cases  the derived RV was strongly  discrepant from the bulk of    
the other stars; this is a real effect, as  confirmed by the spectral    
ranges where atmospheric or interstellar lines     
are present (e.g. the Na {\sc i} D lines).     
We have assumed that those are field objects and have marked them with    
``F'' in Table \ref{tab1}.     
They have not been included in      
the following analysis to avoid introducing spurious effects.      
     
From the 20 stars observed with UVES we derive a mean heliocentric RV of    
100.9 kms$^{-1}$ ($\sigma$ = 8.5), whereas from the lower precision GIRAFFE 
measures we obtain 102.1  kms$^{-1}$ ($\sigma$ = 10.9) using 81 member stars  
observed in the Na {\sc i} region, and 99.0  kms$^{-1}$ ($\sigma$ = 10.3) 
using 83 stars observed in the Ca {\sc ii} setup. 

The first spectroscopic observations of NGC 2808 were taken by T.D. Kinman 
nearly 50 years ago: he set on a bright star in the center and exposed for 
about 3 hours, but because of guiding uncertainties the spectrum obtained 
was at least partially an integrated spectrum (Kinman 2003, private
communication). The mean radial velocity derived from these data was 
101$\pm$5 kms$^{-1}$ (Kinman 1959). Our results are in excellent agreement 
with this earliest determination. 
Later measures of mean radial velocity, summarised by Harris (1996, 
updated 2003) as 93.6$\pm$2.4 kms$^{-1}$, include values ranging 
from e.g.  80.1$\pm$9.9  (Rutledge et al. 1997), to 98$\pm$4 (Hesser et al. 
1986),  to 104.1$\pm$4.4 (Webbink 1981) kms$^{-1}$. 
The most discrepant result, by 
Rutledge et al. (1997), should probably be given little weight, as 
$\sim$25\% of their results on globular clusters differ by more than 
20 kms$^{-1}$ from the average of all previous determinations, and the 
authors say that the goal of their project was not to determine accurate 
cluster velocities. 

Therefore all determinations, excluding  Rutledge et al. (1997), are in 
agreement within 1-$\sigma$ error.

     
\section{The physical and atmospheric parameters}     
     
Assuming for NGC 2808 the reddening E$(B-V)$ = 0.22 and apparent distance      
modulus $(m-M)_V$ = 15.59 (Harris 1996), we have calculated the intrinsic     
colours  $(B-V)_0$ and $(V-K)_0$ and absolute magnitudes of our stars using     
the relations:     
E($V-K$) = 2.75 $\times$ E($B-V$), A$_V$ = 3.1 $\times$ E($B-V$) and      
A$_K$ = 0.35 $\times$ E($B-V$) (Cardelli et al. 1989).      
     
The effective temperatures and bolometric corrections have been obtained      
both from visual and IR colours using two independent empirical calibrations      
specifically derived for late type giant stars by Montegriffo et al.    
(1998, their Table 3) and Alonso et al. (1999, their Eq.s \#4, 9 and 17.    
See also Alonso et al. 2001).      
The metallicity assumed for NGC 2808 is [Fe/H]=--1.25 as an intermediate value    
among several determinations (cf. Walker 1999).        
     
Transformations of the 2MASS K data to the ESO and TCS  photometric systems,      
used by Montegriffo and Alonso respectively,  have been performed using      
the relations provided by Carpenter (2001). The relations we have used are:      
$K_{ESO}=K_{2MASS} - 0.005(J-K)_{2MASS} + 0.045$  and      
$K_{TCS}=K_{2MASS} + 0.006(J-K)_{2MASS} + 0.002$.     
The relation for the TCS photometric system has been obtained via an     
intermediate passage through the CIT system.      
     
The T$_{\rm eff}$'s obtained from these two calibrations are compared in      
Fig. \ref{te_cals}, where we see that they are quite compatible once allowance     
is  made for a systematic offset that makes Alonso temperatures {\em lower} by      
$\simeq$ 153 ($\sigma$=65) and  88 ($\sigma$=17) K, when considering      
T$_{\rm eff}$'s derived from $(B-V)$ and $(V-K)$ respectively.     
     
This difference is irrelevant for the purpose of the      
present paper, where temperatures and luminosities are only used to locate      
correctly the stars in the HR diagram. It might be more relevant for a      
careful estimate of elemental abundances, but this aspect is       
discussed elsewhere (Carretta et al. 2003a,b).      
In the following we shall use Alonso et al. (1999, 2001) calibration as it is      
 widely used and is therefore more convenient for comparison with      
other studies. With this calibration, the $(B-V)$ and $(V-K)$ colours yield      
similar results, except at the ends of the temperature     
range we are interested in, as we can see in Fig. \ref{tes}; the average     
difference between the two T$_{\rm eff}$'s is 19 ($\sigma$=95) K.      
     
We list in Table 2 the values of temperature and luminosity we obtain      
for all our stars from the $(B-V)$ and $(V-K)$ colours and the Alonso et al.      
(1999, 2001) calibration, assuming $M_{{\rm bol},\odot}$ = 4.75.      
The values of gravity       
from the two colours are the same within 0.05 dex in the logarithm, and we      
list only the values derived from the $(B-V)$ colours.

\begin{table*}      
\begin{center}      
\label{t:param}      
\caption[]{Physical and atmospheric parameters for our RGB stars in NGC2808.   
The last three columns give the measured heliocentric radial velocities  
for two GIRAFFE setups (HR02 and 12) and for UVES, derived by averaging  
several photospheric lines.        
}       
\begin{tabular}{rcccccccrrr}      
\hline\hline      
\\      
ID n. & M$_V$ & $(B-V)_0$ & T$_{\rm eff}$ & log L  & log g & T$_{\rm eff}$ &log L &RV$_{02}$ &RV$_{12}$ &RV$_{\rm U}$\\   
      &       &           & $(B-V)$   & $(B-V)$ & $(B-V)$& $(V-K)$& $(V-K)$       &kms$^{-1}$&kms$^{-1}$&kms$^{-1}$  \\   
\\          
\hline      
\\      
  7183 & -0.736 &  1.094&  4432&  2.405	& 1.50&      &         & 100.31&	 &  \\   
  7536&  -1.218&   1.220&  4261&  2.645&  1.19&  4245&  2.650  & 93.65 &  98.83  &\\   
  7558&  -0.201&   0.967&  4618&  2.150&  1.83&  4691&  2.135  &       & 121.32  &\\   
  7788&  -0.720&   1.078&  4454&  2.394&  1.52&  4459&  2.392  & 89.00 & 96.35   &\\   
  8603&  -1.158&   1.195&  4294&  2.612&  1.24&  4290&  2.612  &       & 110.88  &\\   
  8739&  -1.302&   1.253&  4219&  2.692&  1.13&  4211&  2.695  & 110.92& 111.89  & \\   
  8826&  -0.557&   1.022&  4536&  2.310&  1.64&  4564&  2.303  & 95.47 & 98.45   & \\   
  9230&  -1.562&   1.343&  4107&  2.836&  0.94&  4132&  2.827  &       & 89.31   & \\   
  9724&  -1.082&   1.141&  4366&  2.561&  1.32&  4330&  2.570  & 84.83 & 84.33   & \\   
  9992&  -1.888&   1.480&  3949&  3.041&  0.67&  3964&  3.033  & 105.77&	 & \\   
 10012&  -1.365&   1.237&  4239&  2.711&  1.12&  4243&  2.710  & 103.83& 104.26  & \\   
 10105&  -0.921&   1.087&  4442&  2.477&  1.44&  4455&  2.474  & 81.11 &  81.58  & \\   
 10201&   0.124&   0.930&  4676&  2.008&  1.99&  4717&  2.000  &       & 111.21  & 105.96\\   
 10265&  -1.281&   1.199&  4289&  2.662&  1.19&  4283&  2.664  & 106.50& 108.70  & \\   
 10341&  -0.931&   1.130&  4382&  2.496&  1.39&  4399&  2.492  & 99.97 & 102.53  & \\   
 10571&  -1.214&   1.210&  4274&  2.640&  1.21&  4252&  2.646  & 106.20& 112.98  & \\   
 10681&  -2.045&   1.496&  3931&  3.114&  0.59&  4005&  3.074  & 97.07 &	 & \\   
 13575&  -0.204&   1.009&  4555&  2.164&  1.79&  4569&  2.161  &       & 95.33   & \\   
 13983&   0.350&   0.867&  4777&  1.899&  2.14&  4826&  1.890  &       & 110.07  & 109.00\\   
 30523&  -0.101&   0.949&  4646&  2.104&  1.89&  4733&  2.087  &       & 102.23  & \\   
 30763&  -0.984&   1.101&  4422&  2.507&  1.40&  4547&  2.478  &94.10  &	 & \\   
 30900&  -0.506&   1.015&  4546&  2.287&  1.67&  4666&  2.262  & 114.32& 119.94  & \\   
 30927&  -2.074&   1.479&  3950&  3.115&  0.59&  4138&  3.029  & 86.78 &	 & \\   
 31851&  -0.369&   0.979&  4600&  2.220&  1.75&  4731&  2.195  &       &  93.05  & \\   
 32398&  -0.903&   1.097&  4427&  2.473&  1.43&  4494&  2.457  & 94.40 &	 & \\   
 32469&  -1.550&   1.225&  4255&  2.780&  1.06&  4379&  2.744  &       &  94.15  & \\   
 32685&   0.066&   0.918&  4695&  2.028&  1.98&  4788&  2.010  &       &  87.17  &  94.52\\   
 32862&  -0.424&   0.993&  4579&  2.247&  1.72&  4525&  2.258  & 98.57 & 105.16  & \\   
 32909&  -0.568&   1.030&  4524&  2.316&  1.63&  4572&  2.306  &       &  90.43  & \\   
 32924&  -0.737&   1.065&  4473&  2.396&  1.53&  4538&  2.381  &       & 101.03  &\\   
 33452&  -0.420&   1.015&  4546&  2.252&  1.70&  4671&  2.226  & 92.79 &	& \\   
 33918&  -1.259&   1.225&  4255&  2.664&  1.17&  4312&  2.647  & 117.58& 119.80  &\\   
 34008&  -0.471&   1.007&  4558&  2.270&  1.69&  4670&  2.247  & 102.73&	 &\\   
 34013&   0.854&   0.835&  4830&  1.688&  2.37&  5110&  1.645  &       &	 &101.49 \\   
 34634&  -0.229&   1.013&  4549&  2.175&  1.78&  4627&  2.159  &       &	 &\\   
 35265&  -0.269&   1.043&  4505&  2.201&  1.74&  4547&  2.192  &       &  98.95  &\\   
 37496&  -0.719&   1.035&  4516&  2.378&  1.56&  4547&  2.372  & 109.69& 109.75  &\\   
 37505&  -0.807&   1.055&  4487&  2.420&  1.51&  4504&  2.416  & 96.87 &	& \\   
 37872&  -1.940&   1.464&  3967&  3.052&  0.66&  4015&  3.028  & 104.96&	 & 104.94\\   
 37998&  -1.357&   1.185&  4307&  2.687&  1.17&  4317&  2.684  & 90.72 &  94.27  &\\   
 38228&  -0.609&   1.010&  4553&  2.326&  1.63&  4588&  2.319  &       & 101.96  &\\   
 38244&  -0.397&   0.965&  4621&  2.227&  1.75&  4659&  2.220  &       & 106.87  &\\   
 38660&  -1.300&   1.162&  4338&  2.656&  1.22&  4332&  2.657  & 96.62 &	 &\\   
 38967&  -0.338&   0.943&  4655&  2.197&  1.80&  4717&  2.185  &       &  98.44  & \\   
 39060&  -1.173&   1.156&  4346&  2.602&  1.27&  4357&  2.600  &       & 113.95  & \\   
 40196&  -1.020&   1.104&  4418&  2.522&  1.38&  4413&  2.524  & 97.76 &	& \\   
 40983&  -1.230&   1.143&  4364&  2.620&  1.26&  4364&  2.620  & 106.19& 118.90  &\\   
 41008&  -0.401&   0.930&  4676&  2.218&  1.78&  4684&  2.216  &       &  94.62  &\\   
 41828&  -1.069&   1.121&  4394&  2.548&  1.35&  4409&  2.544  &       & 100.96  &\\   
 41969&  -1.144&   1.121&  4394&  2.578&  1.32&  4377&  2.583  & 79.00 & 88.81  & \\   
 42073&  -1.299&   1.191&  4299&  2.666&  1.19&  4268&  2.676  &       & 101.15  &\\   
 42165&   0.797&   0.823&  4851&  1.707&  2.36&  4906&  1.698  &       & 108.81  &\\   
 42789&  -0.223&   0.889&  4741&  2.134&  1.89&  4813&  2.122  &       & 115.68  &\\   
 42886&   0.331&   0.878&  4759&  1.910&  2.12&  4791&  1.904  &       & 109.36  & 114.20\\   
 42996&  -0.164&   0.906&  4714&  2.116&  1.90&  4769&  2.106  &       &  91.47  &\\   
 43041&  -1.867&   1.406&  4033&  2.990&  0.75&  4035&  2.989  & 112.22& 109.48  &\\   
   \\     
\hline      
\end{tabular}      
\end{center}      
\end{table*}      
     
\addtocounter{table}{-1}        
     
\begin{table*}      
\begin{center}      
\caption[]{(Continuation)       
}       
\begin{tabular}{rcccccccrrr}      
\hline\hline      
\\      
ID n. & M$_V$ & $(B-V)_0$ & T$_{\rm eff}$ & log L  & log g & T$_{\rm eff}$ &log L &RV$_{02}$ &RV$_{12}$ &RV$_{\rm U}$ \\   
      &       &           & $(B-V)$   & $(B-V)$ & $(B-V)$& $(V-K)$& $(V-K)$       &kms$^{-1}$&kms$^{-1}$&kms$^{-1}$   \\   
\\      
\hline      
\\      
 43217&   0.850&   0.811&  4871&  1.683&  2.39&  4916&  1.676 &        &	& 103.23\\   
 43247&  -0.804&   1.016&  4544&  2.406&  1.55&  4463&  2.425 & 106.76 & 107.76 &\\   
 43281&  -0.543&   1.003&  4564&  2.298&  1.66&      &        &  95.23 &	&\\   
 43561&  -2.267&   1.619&  3801&  3.294&  0.35&  3872&  3.240 & 92.69  &        & \\   
 43794&  -1.036&   1.087&  4442&  2.523&  1.39&  4431&  2.526 & 97.95  & 101.91 & \\   
 43822&  -0.990&   1.111&  4408&  2.513&  1.39&  4411&  2.512 & 88.92  & 99.96  &\\   
 44573&  -0.501&   0.976&  4605&  2.272&  1.70&  4516&  2.291 & 88.63  &        & \\   
 44665&  -1.345&   1.184&  4308&  2.682&  1.18&  4310&  2.681 & 91.94  & 93.67  & \\   
 44716&  -0.691&   1.004&  4562&  2.357&  1.60&  4539&  2.362 & 99.30  &        & \\   
 44984&  -1.055&   1.100&  4423&  2.535&  1.37&  4405&  2.540 & 86.65  & 82.96  &\\   
 45162&  -2.327&   1.689&  3731&  3.386&  0.22&  3810&  3.310 & 112.90 &	&\\   
 45443&  -1.016&   1.091&  4436&  2.516&  1.39&  4425&  2.519 & 99.23  & 96.10  & \\   
 45840&  -1.473&   1.215&  4268&  2.745&  1.10&  4220&  2.760 & 95.49  & 100.90 &\\   
 46041&  -0.623&   0.979&  4600&  2.322&  1.65&  4568&  2.329 & 93.63  & 103.22 & \\   
 46099&  -1.849&   1.414&  4024&  2.987&  0.75&  4032&  2.983 & 102.55 &	& 103.48\\   
 46367&  -0.758&   1.002&  4565&  2.383&  1.58&  4383&  2.427 &        & 95.19  &\\   
 46422&  -2.214&   1.578&  3843&  3.239&  0.42&      &        &        &	& 93.50 \\   
 46580&  -1.900&   1.382&  4061&  2.991&  0.77&      &        & 102.34 &	& 103.12 \\   
 46663&  -1.039&   1.089&  4439&  2.525&  1.39&  4474&  2.516 &        & 97.26  &\\   
 46726&  -1.215&   1.145&  4361&  2.615&  1.27&  4360&  2.616 &  90.03 &	&\\   
 46868&  -0.838&   1.031&  4522&  2.425&  1.52&      &        &        & 110.38 &\\   
 46924&  -0.764&   1.050&  4495&  2.402&  1.53&      &        & 96.36  &	& \\   
 47031&  -1.409&   1.204&  4282&  2.715&  1.13&  4076&  2.788 & 95.43  &	&\\   
 47145&  -1.943&   1.382&  4061&  3.008&  0.75&  4051&  3.012 & 66.67  & 75.97  &\\   
 47421&  -1.415&   1.176&  4319&  2.707&  1.16&  4288&  2.716 &  74.65 & 69.43  &\\   
 47452&  -1.813&   1.860&  3570&  3.457&  0.08&  3737&  3.174 & 101.28 &	&\\   
 47606&  -2.154&   1.601&  3819&  3.233&  0.42&  3839&  3.218 &  96.25 &	& 97.77\\   
 48011&  -1.193&   1.101&  4422&  2.591&  1.31&  4375&  2.603 &  98.87 & 101.85 &\\   
 48060&  -2.034&   1.480&  3949&  3.099&  0.61&  3969&  3.088 &  96.03 & 91.58  &\\   
 48128&  -1.106&   1.121&  4394&  2.563&  1.33&  4368&  2.570 &  96.16 & 101.22 &\\   
 48424&  -1.138&   1.100&  4423&  2.568&  1.34&  3649&  3.020 & 109.01 &	&\\   
 48609&  -2.175&   1.650&  3769&  3.286&  0.34&  3846&  3.221 &        &	&  85.20\\   
 48889&  -2.249&   1.578&  3843&  3.253&  0.41&  3943&  3.188 & 114.00 &	& 116.29\\   
 49509&  -1.669&   1.288&  4175&  2.854&  0.95&  4164&  2.858 &        & 106.58 &\\   
 49680&  -2.023&   1.469&  3961&  3.088&  0.63&  3952&  3.094 &        & 83.64  &\\   
 49743&  -0.422&   0.965&  4621&  2.237&  1.74&  4665&  2.228 & 107.18 & 112.48 &\\   
 49753&  -1.240&   1.099&  4425&  2.609&  1.30&  4298&  2.643 & 112.04 &	&\\   
 49942&  -1.854&   1.444&  3989&  3.006&  0.72&  4023&  2.990 &        & 118.85 &\\   
 50119&  -1.704&   1.319&  4136&  2.882&  0.91&  4166&  2.871 & 100.83 &	& 101.56 \\   
 50371&  -1.750&   1.404&  4035&  2.942&  0.80&  4054&  2.934 &        & 104.36 &\\   
 50561&  -0.756&   1.066&  4471&  2.404&  1.52&  4429&  2.414 &  92.71 &	&\\   
 50681&  -1.930&   1.423&  4013&  3.025&  0.71&  4114&  2.981 & 115.24 &	&\\   
 50761&  -2.200&   1.784&  3639&  3.462&  0.10&  3756&  3.309 & 103.34 &	& 99.80 \\   
 50861&  -2.034&   1.425&  4011&  3.067&  0.67&  4039&  3.054 & 108.07 & 110.60 &\\   
 50866&  -0.956&   1.087&  4442&  2.491&  1.42&  4432&  2.493 &        & 95.50  &\\   
 50910&  -0.606&   1.030&  4524&  2.332&  1.61&  4577&  2.320 &        & 98.33  &\\   
 51416&  -0.645&   1.023&  4534&  2.345&  1.60&  4546&  2.342 & 87.40  & 90.40  &\\   
 51454&  -2.144&   1.567&  3855&  3.202&  0.46&  3893&  3.177 &  82.79 &	& 83.35\\   
 51499&  -2.155&   1.502&  3925&  3.162&  0.54&  3960&  3.142 & 108.71 &	& 109.18\\   
 51515&  -1.377&   1.379&  4065&  2.780&  0.98&  4154&  2.745 & 104.52 & 106.77 &\\   
 51646&  -0.130&   0.870&  4772&  2.092&  1.95&  4886&  2.072 &        &  99.69 & \\   
 51871&  -2.048&   1.547&  3876&  3.149&  0.53&  3970&  3.094 &        &  93.84 &\\   
 51930&  -1.522&   1.243&  4232&  2.776&  1.05&  4214&  2.782 &  90.34 &	&\\   
 51963&  -1.391&   1.233&  4244&  2.720&  1.11&  4242&  2.720 & 100.59 & 102.54 &\\   
 51983&  -2.116&   1.634&  3786&  3.247&  0.39&  3855&  3.191 &        &	& 101.48 \\   
 52006&  -1.268&   1.111&  4408&  2.624&  1.27&  4397&  2.627 & 108.92 & 109.48 &\\   
   \\     
\hline      
\end{tabular}      
\end{center}      
\end{table*}      
     
\addtocounter{table}{-1}        
     
\begin{table*}      
\begin{center}      
\caption[]{(Continuation)       
}       
\begin{tabular}{rcccccccrrr}      
\hline\hline      
\\      
ID n. & M$_V$ & $(B-V)_0$ & T$_{\rm eff}$ & log L  & log g & T$_{\rm eff}$ &log L &RV$_{02}$ &RV$_{12}$ &RV$_{\rm U}$ \\   
      &       &           & $(B-V)$   & $(B-V)$ & $(B-V)$& $(V-K)$& $(V-K)$       &kms$^{-1}$&kms$^{-1}$&kms$^{-1}$   \\   
\\      
\hline      
\\      
 52048&  -1.331&   1.166&  4333&  2.670&  1.20&  4294&  2.681 & 75.18  &         &\\   
 52647&  -0.936&   1.051&  4493&  2.471&  1.46&  4480&  2.474 & 121.06 &         &\\   
 53284&  -1.743&   1.434&  4001&  2.956&  0.77&  4038&  2.938 & 104.62 &         &\\   
 53390&  -0.921&   1.069&  4467&  2.471&  1.45&  4426&  2.481 &        &         & 99.71 \\   
 53579&  -1.129&   1.149&  4356&  2.582&  1.30&  4294&  2.600 &  94.66 &         &\\   
 54264&  -0.254&   0.919&  4693&  2.156&  1.85&  4668&  2.161 &        & 121.39  &\\   
 54284&  -1.183&   1.162&  4338&  2.609&  1.26&  4327&  2.612 &        & 117.17  &\\   
 54308&  -1.931&   1.689&  3731&  3.228&  0.38&  3732&  3.226 & 109.80 &         &\\   
 54733&  -1.462&   1.254&  4217&  2.757&  1.07&  4236&  2.751 & 111.60 & 113.25  &\\   
 54756&  -0.806&   1.082&  4449&  2.429&  1.49&  4452&  2.428 &  94.31 &         &\\   
 54789&  -1.343&   1.159&  4342&  2.672&  1.20&  4355&  2.668 &  94.68 &  88.05  &\\   
 55031&  -1.990&   1.463&  3968&  3.072&  0.64&  3995&  3.058 &        & 106.71  &\\   
 55354&  -0.841&   1.055&  4487&  2.434&  1.50&  4483&  2.435 & 109.67 &         &\\   
 55437&  -0.898&   1.070&  4466&  2.462&  1.46&  4440&  2.468 &        & 106.19  &\\   
 55609&  -0.839&   1.107&  4413&  2.451&  1.45&  4430&  2.447 & 114.62 & 118.14  &\\   
 55627&  -1.005&   1.094&  4432&  2.513&  1.40&  4459&  2.506 &        &  96.45  &\\   
 56032&  -1.614&   1.383&  4060&  2.877&  0.88&  4045&  2.884 &        &         & 90.42 \\   
 56136&   0.793&   0.805&  4882&  1.704&  2.37&  4905&  1.700 &        & 115.11  &\\   
 56536&  -1.964&   1.455&  3977&  3.056&  0.66&  4016&  3.037 & 107.17 &         &\\   
 56710&  -0.160&   0.932&  4673&  2.122&  1.88&  4643&  2.128 &        & 115.35  &\\   
 56924&  -1.383&   1.202&  4285&  2.704&  1.15&  4252&  2.714 &  98.65 &         &\\   
   \\     
\hline      
\end{tabular}      
\end{center}      
\end{table*}      
       
\section{Results and discussion}      
     
\subsection{H$\alpha$ line}      
     
\begin{figure*}      
\begin{center}     
\includegraphics[width=18cm]{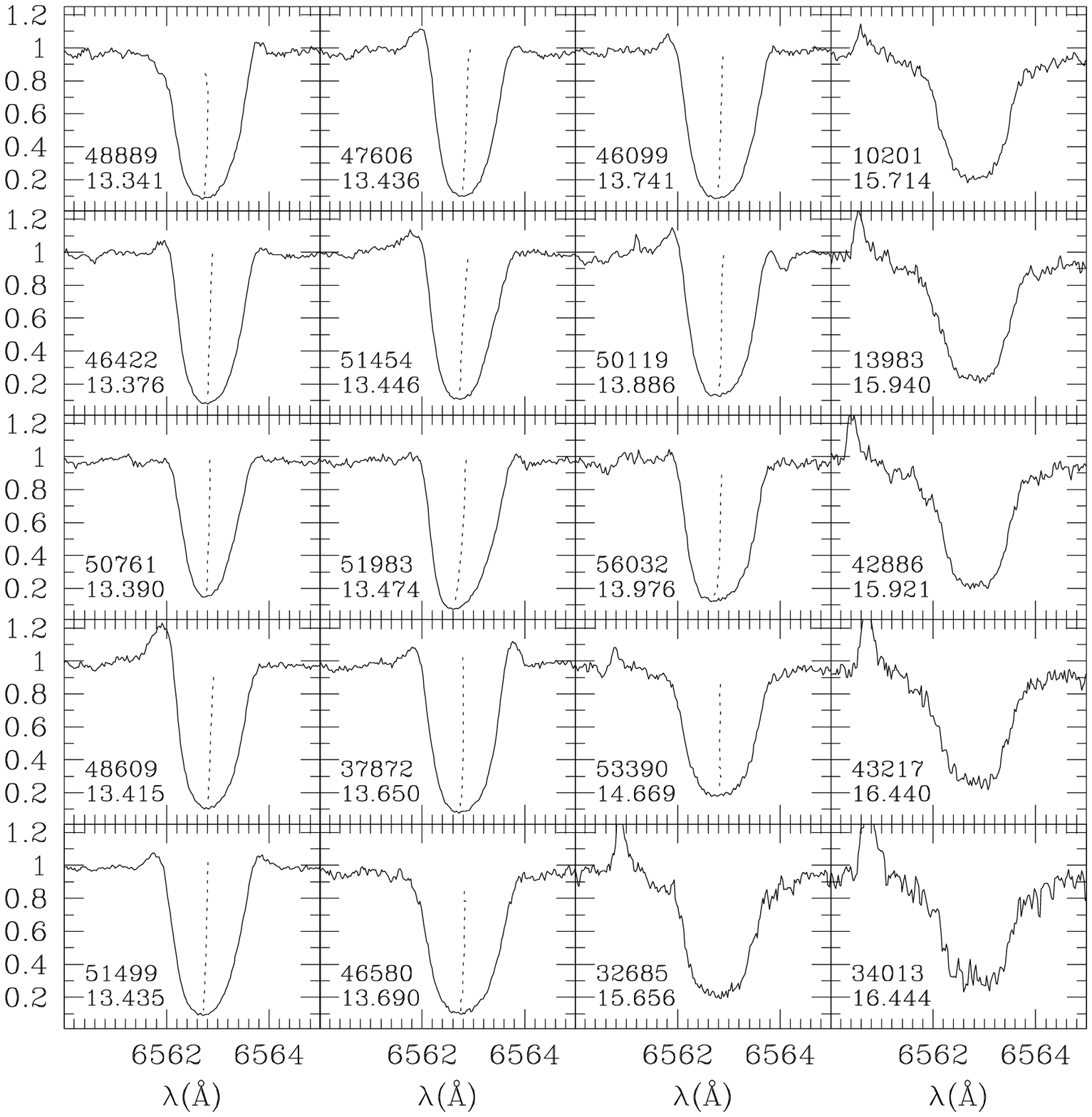}       
\caption{Normalised parts of spectra containing the H$\alpha$ line for the 20      
stars observed with UVES, shifted to RV=0.  
We also plot the line bisectors for the brighter stars. 
Note that spectra for stars fainter than V=14.5  
show the sky H$\alpha$ line in emission on the blue side of the stellar  
H$\alpha$ line. }      
\label{halpha}      
\end{center}     
\end{figure*}

\subsubsection{Emission}     
     
\begin{figure*}      
\begin{center}     
\includegraphics[width=18cm]{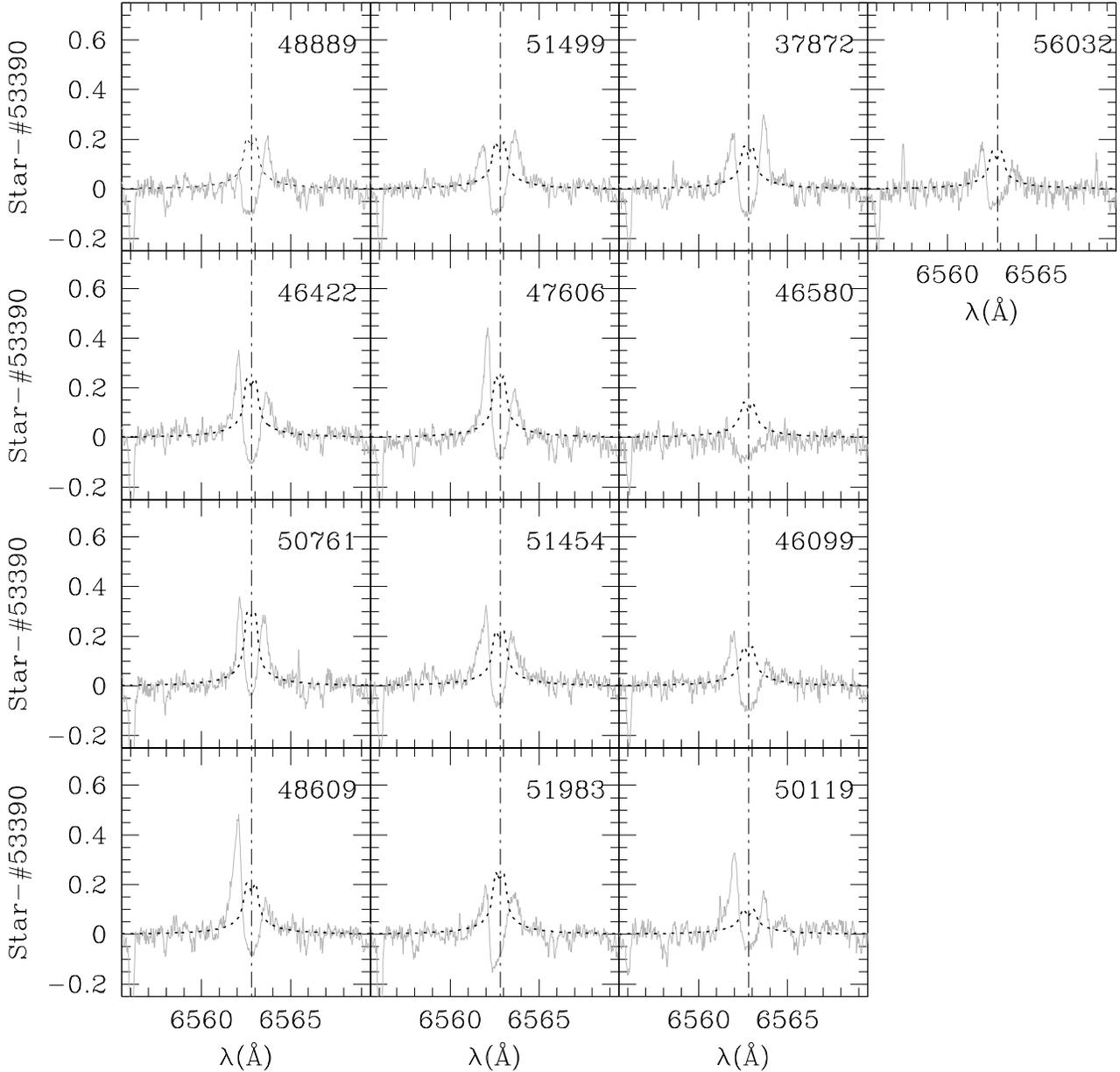}       
\caption{H$\alpha$ lines after subtraction of the template profile from star    
\#53390, for the 13 stars observed with UVES and brighter than V=14.0    
($\log L/L_{\odot}$=2.88). The solid grey lines show the differences of 
the observed profiles, the dotted lines show the differences of the 
corresponding theoretical profiles.}      
\label{hasub}      
\end{center}     
\end{figure*}

We show in Fig. \ref{halpha} the H$\alpha$ line profiles for the 20 stars observed with      
UVES. One can see that most of the stars brighter than V=14 show clear    
evidence of emission wings, mostly on the blue side, but also on both sides,    
whereas the fainter ones do not show such features. We note that weak emission    
from the wings could escape detection as it would be compensated by the    
absorption, the net result being a narrower absorption profile. In order to    
reveal all possible evidence of emission, after careful visual inspection     
we have arbitrarily identified a ``template'' star with no obvious H$\alpha$    
emission and a fairly symmetric profile.      
This star is \#53390 (V=14.67, $M_V$=--0.9, $T_{\rm eff}\sim4400$K   
and $\log L/L_{\odot}\sim$2.5). We have then subtracted the H$\alpha$ profile    
of \#53390 from all stars in our UVES sample brighter than V=14 after    
normalizing the intensities to the continuum and shifting in wavelength to    
superpose the bisector of the H$\alpha$ profile at the level of half maximum    
intensity.    
We have applied this procedure only to the stars in our UVES sample brighter    
than V=14.5 because the fainter ones become progressively    
hotter and the H$\alpha$ profiles have a different shape, besides showing    
no indication whatsoever of emission wings. 
However, the possible dependence of the H$\alpha$ line shape on 
temperature might introduce spurious features in the subtracted profiles. 
In order to test for this effect, we have calculated the theoretical 
H$\alpha$ profiles for our stars using Kurucz model atmospheres (with no 
chromosphere)  and the individual values of temperature and gravity estimated 
from the (B--V) colors that are listed in Table 2. 
In Fig. \ref{hasub} we show the difference of the observed spectra 
(star--\#53390) 
and, for comparison, the difference of the corresponding theoretical profiles. 
Clearly the observed profiles are quite different from the theoretical ones as 
predicted just from the variations of temperature and gravity along the RGB. 
In particular: 
i) In all stars the absorption at the center of the line is larger than in 
the template star, contrarily to the theoretical expectations. This suggests 
the presence of a thicker atmosphere (i.e. an excess of material) and/or 
a higher temperature in the external regions (i.e. a chromosphere). 
This absorption is often blue-shifted, as it clearly appears in the stars 
48889, 51499, 51454, 51983, 37872, 46580 and 56032. 
ii) Whereas the red emission wing could be due to the subtraction procedure 
in some case (see e.g. the stars 48609 or 56032), the blue emission wing 
seems real in all cases (except perhaps for star 51983). We note that the 
stars 48889 and 46580, that do not show blue emission, do however show an 
asymmetric blue-shifted absorption core like star 51983. 
From Fig. \ref{hasub}, therefore,  it appears that 10 out of 11 stars 
brighter than $\log L/L_{\odot}\sim$2.9 (i.e. more than 90\%)    
show H$\alpha$ emission wings.   
The faintest stars of this group, \#50119 and 56032, have    
$\log L/L_{\odot}$=2.87-2.88; however we cannot take this value as a threshold    
for the onset of emission wings, since there is a gap of about half a    
magnitude in our UVES sample and the monitoring is not continuos in luminosity. 
   
We have applied a similar procedure to the stars with $\log L/L_{\odot}>2.5$   
that were observed with GIRAFFE/MEDUSA and setup HR14.  
Unlike for the UVES stars, we did not attempt to estimate     
the velocities of the emission peaks with respect to the bisector of the    
H$\alpha$ absorption profile, because of the shaky wavelength calibration. 
Instead, we have shifted spectra to 
RV=0 using nearby photospheric lines.  In this case the    
template spectrum used for subtraction is the average of three spectra, i.e.    
those of stars \#8603, 10571 and 43822, that have temperatures in the range    
$T_{\rm eff}$=4250-4400K and luminosities in the range $\log L/L_{\odot}$=2.50-2.65.    
These parameters are quite similar  to those of star    
\#53390 that was selected as a template for the UVES spectra.    
The much larger number of stars observed with GIRAFFE gives us a more    
detailed monitoring of the H$\alpha$ line behaviour as a function of    
luminosity, but of course the GIRAFFE lower resolution might more easily    
hide weak  H$\alpha$ emission wings in the strong H$\alpha$ absorption line.    
   
From the GIRAFFE sample we estimate that the fractions of stars showing    
some evidence of H$\alpha$ emission are approximately as indicated in   
parenthesis, in the luminosity intervals $\log L/L_{\odot}$=2.5-2.6 (30\%),    
2.6-2.7 (73\%), 2.7-2.8 (86\%), 2.8-2.9 (100\%) and $>$2.9 (100\%). If we   
include also the UVES results, in the interval $\log L/L_{\odot}>$2.9    
the emission detection frequency would be $\sim$ 95\%,  and $\sim$ 94\%    
for $\log L/L_{\odot}>$2.7.

A comparison with previous studies  must take into account that the various    
sets of results were obtained with different procedures and with spectra of    
different resolution.     Our procedure of subtracting an H$\alpha$ template, 
i.e. a supposedly pure  absorption    profile, was not chosen by anybody else 
before. In principle it should  allow    to detect emission features that might 
have gone unnoticed had this subtraction not been performed, therefore we 
would expect a higher detection rate because of this.     
On the other hand, we compare our UVES and GIRAFFE data, with spectral    
resolution of $\sim$7 and 10 km$^{-1}$ respectively, with the major statistical    
studies on H$\alpha$ emission in metal-poor red giant stars, namely    
Cacciari \& Freeman (1983, 143 RGB stars in globular clusters,    
res. $\sim$30 kms$^{-1}$), Gratton et al. (1984, 113 RGB stars in globular   
clusters, res. $\sim$ 15 kms$^{-1}$), Smith \& Dupree (1988, 52 metal-poor field    
red giants, res. $\sim$7 kms$^{-1}$), and Lyons et al. (1996, 62 RGB stars    
in globular clusters, res. $\sim$5 kms$^{-1}$). A poorer resolution    
makes the detection of emission features more difficult.        
As discussed also by Lyons et al. (1996), H$\alpha$ emission was detected    
by Cacciari \& Freeman (1983) only among stars brighter than    
$\log L/L_{\odot}=$2.9 in a proportion of $\sim$33\%, by Gratton et al. (1984)   
among stars brighter than $\log L/L_{\odot}=$2.7 in a proportion of $\sim$61\%,   
by Smith \& Dupree (1988) above the threshold of $M_V=-1.7$    
(i.e. $\log L/L_{\odot}\sim2.9$) in a proportion of $\sim$77\% ($\le$50\% if    
reported to the threshold of $\log L/L_{\odot}=$2.7), and finally by    
Lyons et al. (1996) above the threshold of $\log L/L_{\odot}=$2.7 in a    
proportion of $\sim$ 80\%.    
   
Our detection limit for H$\alpha$ emission is $\log L/L_{\odot}\sim$2.9    
from the UVES stars with a proportion of $\sim$91\%;       
from the GIRAFFE stars our detection threshold is $\log L/L_{\odot}\sim$2.5    
with a proportion of $\sim$72\%.  If we report these results to the threshold    
value of $\log L/L_{\odot}=$2.7 for the sake of comparison, and consider both    
UVES and GIRAFFE stars, then $\sim$94\% of the stars brighter than this value    
exhibit H$\alpha$ emission wings.    
     
Finally, we note that  time variability of H$\alpha$ emission would contribute 
to underestimate  the frequency of H$\alpha$ emitting stars. 
The present set of data, however,      
does not allow us to monitor for possible variability of H$\alpha$ emission       
features.      
     
\subsubsection{H$\alpha$ emission: mass loss or chromospheric activity?}

The nature of the H$\alpha$ emission is controversial. Early studies (cf. 
Cohen 1976; Mallia \& Pagel 1978; Cacciari \& Freeman 1983; Gratton et al. 
1984) assumed that it could be taken as evidence of extended matter around the 
star hence mass loss could be inferred. However, Dupree et al. (1984) argued 
that H$\alpha$ emission could form naturally in a static chromosphere, or it 
could be enhanced by hydrodynamic processes related to pulsation (Dupree et 
al. 1994), with no  need to invoke mass loss (cf. also Reimers 1981). We 
discuss here briefly these two hypotheses. 

In the hypothesis that the 
H$\alpha$ emission is due to circumstellar material, one could compare the 
observed EW of the emission components with ``theoretical'' predictions. These 
predictions are obtained by the use of two different relations, one by Cohen 
(1976) that gives the mass loss rate as:\\ 
$\dot{M}=2.4\times10^{-11}V_{exp}R_{\ast}(R_sW_{\lambda})^{1/2}\times 
e^{(-1.1/T_4)}~M_{\odot}yr^{-1}$\\ 
where it is assumed that the emission forms in a shell at constant expansion 
velocity $V_{exp}$ (taken as the observed velocity of the blue peaks listed in 
Table 3) and located at a distance ($R_s$) of 2 stellar radii ($R_{\ast}$) 
from the star, $W_{\lambda}$ is the equivalent width in \AA~ of the emission 
components, and $T_4$ is the brightness temperature of the star in units of 
10$^4$ K. \\ 
The other relation, by Reimers (1975), gives the mass loss rate as:\\ 
$\dot{M}=-4\times10^{-13}\eta L/gR$\\ 
where all quantities are in solar units and $\eta$ is an empirical 
scaling factor that can reproduce reasonably well the HB morphology of 
globular clusters if allowed to vary in the range from 0.25 to 1. 
 
\begin{figure}      
\begin{center}     
\includegraphics[width=9cm]{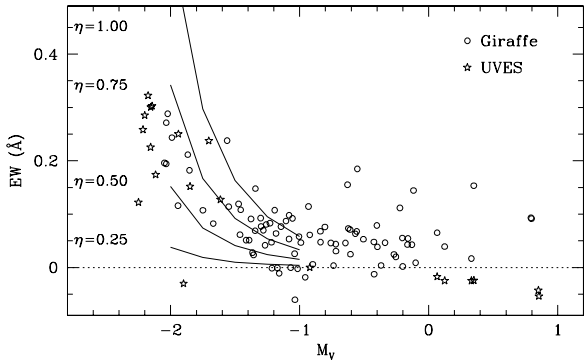}       
\caption{EWs of the emissions on the H$\alpha$ line wings, as a function of  
$M_V$. The lines show what is expected for different values of $\eta$.}      
\label{hareimers}      
\end{center}     
\end{figure}

By imposing that these two parameterizations yield the same value of mass loss 
rate, and using parameter values that can be derived from the observations or 
assumed under reasonable assumptions, one can then derive ``theoretical'' 
values for the EW's of the emission components, to be compared with the 
observed ones. Fig. \ref{hareimers} shows a plot of the EW of the emission 
components, both for stars observed with UVES (starred symbols) and with 
GIRAFFE (open circles).  A good fit of the estimated and observed EWs can be 
obtained for a value of $\eta\sim$0.5. This result does not mean that we are 
indeed observing mass loss via the H$\alpha$ emission wings, but only that 
this possibility is consistent with the admittedly very approximate and 
uncertain parameterizations available so far. 
 
\begin{figure}      
\begin{center}     
\includegraphics[width=9cm]{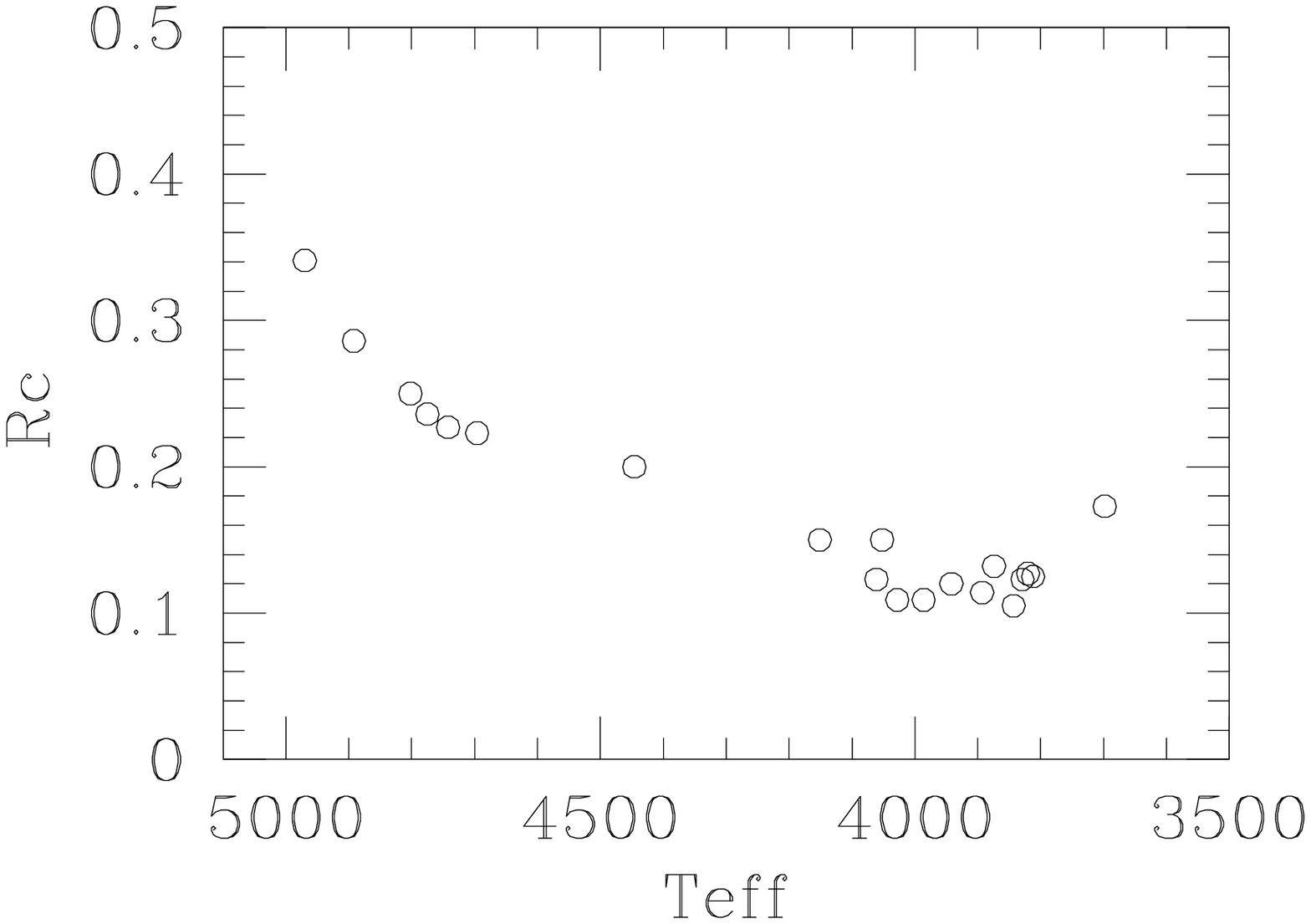}       
\caption{Central residual intensity Rc  of the H$\alpha$ line as a 
function of temperature, for the UVES stars only.  }      
\label{rc}      
\end{center}     
\end{figure}

A difficulty with the Cohen circumstellar model, where the emission region 
is detached from the stellar atmosphere, is that it does not predict 
correctly the residual intensity at the center of the H$\alpha$\ absorption. 
According to this model, in fact, the residual intensity should be larger 
(if some emission occurs at the stellar radial velocity), or at least equal 
to the undisturbed photospheric absorption. Therefore, we would expect that 
H$\alpha$\ should be weaker in cooler stars and  the residual 
intensity at the line center should increase with decreasing temperature. 
However, the opposite holds, as shown in Fig. \ref{rc}.  
The strong absorption at the center of H$\alpha$\ could be justified by  
assuming that the circumstellar material is disposed like a rotating torus  
around the star, but this explanation appears unjustified for single 
stars and it does not support mass loss. 
 
The other hypothesis we are investigating is that the H$\alpha$\ emission is 
due to chromospheric activity. 
The effects of a chromosphere on the profile of H$\alpha$\ depend 
on the amount of material in the chromosphere itself. When the chromosphere is 
optically thin, the emission shows up at the center of the line, as it is the 
case for active (Population I) subgiants (Pasquini \& Pallavicini 1991). In 
this case the emission causes a filling of the core of the line - its 
width is almost unaffected being essentially determined by thermal broadening. 
However, when the chromosphere is optically thick at the center of H$\alpha$\ 
(this may occur if the lower level of the transition is populated by 
recombination from a strong enough UV flux in an extended and dense 
chromosphere), the line profile becomes very different. The mean free path of 
photons at the H$\alpha$\ wavelength becomes short, and they cannot escape the 
atmosphere, unless they are slightly shifted in wavelength by anelastic 
diffusion processes: in this case, they will exit the atmosphere, causing 
blue- or red-shifted emissions. Detailed calculations of models with extended 
atmospheres by Dupree et al. (1984) indeed show  deep central H$\alpha$ 
absorption with residual intensities $\le$0.1, as well as  blue- and 
red-shifted emissions. These predictions agree very well with our results 
for the coolest stars in our sample. We think this gives a strong support 
to the chromospheric explanation of the H$\alpha$\ emission. 
 
It should be reminded that the presence of an extended chromosphere 
neither excludes nor requires mass loss. 
This is shown by the wind model by Dupree et al. (1984), that describes a star 
with an extended chromosphere and a mass loss of 
$2\times 10^{-9}$~M$\odot\,yr^{-1}$. In the case of a net outflow motion, 
slightly red-shifted photons have a higher probability of escape, 
hence we expect that the red-shifted emission will be stronger than the 
blue-shifted one (see Fig.~3 of Dupree et al. 1984). 
We have estimated the relative strengths of the blue (B) and red (R) emission 
wings in our stars by integrating the flux distribution on the difference 
spectra  in two 1\AA-wide bands on the blue and red side of the absorption 
line  (i.e. $\lambda\lambda$6561.3-6562.3 and 6563.3-6564.3), and further 
checked by  visual inspection. The parameter B/R is reported in Tables 3 and 
4 for the stars observed with UVES and GIRAFFE respectively, as    
$<$1 if the red emission wing is stronger (denoted as red asymmetry),    
$>$1 if the blue emission wing is stronger (blue asymmetry),    
and $\sim$1 if the red and blue wings are of similar strength: blue 
asymmetry appears to dominate.  This confirms and reinforces the  
observational evidence found by Smith \& Dupree (1988) among  
metal-deficient field red giants.   
Since all previous studies that could use repeated observations of the same 
stars found that both the emission-line strengths and  
the sense of the B/R asymmetry may change with time (on a timescale of  
months or even days) for any given star, no firm conclusions can be drawn 
from our data. We can only suggest the possible presence of differential mass 
motions in the  line-forming region. 
Smith \& Dupree (1988) proposed an alternative  
explanation of the variability of the emission strength  
(that we cannot detect but may indeed be present in our stars as well),  
i.e. fluctuations of the column density or the temperature gradient or  
both, within the chromosphere, possibly induced by pulsation.  

In conclusion, as already noted by Dupree et al. (1984), asymmetries can be 
altered by episodic events, that may well occur in a (likely) variable 
chromosphere, as well as from a more complex geometry. 
Much more information should be obtained by examining the absorption profile, 
looking for evidence of core blue-shifts. These will be discussed in the next 
subsection.

\begin{table}      
\begin{center}      
\label{t:abu1}      
\caption[]{Parameters of the H$\alpha$ line for the stars observed with UVES.    
Columns 2-3 refer to the absorption core, columns 4-8 refer to the emission 
features.    
The parameter B/R is the intensity ratio of the Blue and Red emission wings    
after subtraction of the template spectrum.     
}      
\begin{tabular}{rrrrrrrr}      
\hline\hline      
\\      
ID n.&Centre &Core  &B/R &Blue &Blue  &Red  &Red  \\	  
     &Sh.    &Sh. & 	 &Pk.  &Ter.  &Pk.  &Ter. \\	  
\\      
\hline      
\\             
   10201   &-1.20 & -0.76&&&&& \\       
   13983   &-0.31 &  0.60&&&&& \\	 
   32685   & 1.07 & -0.76&&&&& \\       
   34013   & 2.43 &  2.43&&&&& \\       
   37872   &-0.76 & -2.57& $<$1 & -44.3 &-79.3 & 39.8 & 57.7\\	  
   42886   & 0.60 &  0.60&&&&& \\       
   43217   &-0.31 &  1.99&&&&& \\	      
   46099   & 1.52 & -1.20& $>$1 &-42.3 &-68.8 & 49.9 & 66.4 \\	  
   46422   & 1.52 & -0.76& $>$1 &-35.7 & -77.5 & 41.1 & 79.2 \\	  
   46580   & 0.16 & -2.57&&&&& \\	 
   47606   & 2.43 & -0.31& $>$1 &-34.6 & -92.2 & 38.3 & 66.4 \\	  
   48609   & 1.99 & -0.31& $>$1 &-39.1 & -86.7 & 38.7 & 65.5 \\	 
   48889   &-0.76 & -3.48& $<$1 &-33.9 & -42.4 & 39.6 & 61.8 \\	  
   50119   & 1.99 & -0.76& $>$1 &-39.6 & -86.7 & 39.3 & 57.3 \\	   
   50761   & 0.60 & -1.20& $\sim$1 &-30.9 & -67.0 & 31.9 & 52.8\\	   
   51454   &-1.20 & -3.95& $>$1 &-43.3 & -83.0 & 35.3 & 75.5 \\	   
   51499   &-1.65 & -4.39& $<$1 &-47.4 & -85.7 & 39.3 & 79.2 \\	   
   51983   &-1.20 & -8.05& $\sim$1 & -41.6 &-83.0 & 35.2 & 55.9\\	 
   53390   & 0.60 & -0.76&&&&& \\   
   56032   & 0.60 & -4.39& $>$1 &-38.5 & -60.7 & 38.3 & 50.8 \\	 
\\      
\hline      
\end{tabular}      
\end{center}      
     
Notes: \\   
i) All radial velocities (centre and core shifts,  and velocities    
of emission peaks and terminal emission profiles) are given in kms$^{-1}$.    
Typical errors of individual measures are $\pm 1kms^{-1}$.  \\   
ii)  The centre and core shifts are relative to the rest position of 
H$\alpha$, while the velocities of the emission peaks and terminal emission 
profiles are relative to the H$\alpha$ line centre (bisector at half maximum).      
     
\end{table}

\begin{table}      
\begin{center}      
\label{t:abu1}      
\caption[]{Stars observed with  GIRAFFE and showing H$\alpha$ emission.  The 
parameter B/R indicates the relative strength of the H$\alpha$  blue and red 
emission wings after subtracting the template spectrum.    
}      
\begin{tabular}{rcl}      
\hline\hline      
\\      
ID n.& B/R &Comments   \\         
\\      
\hline      
\\     
 7536 &  $\sim$1 & \\          
 8739 &  $\sim$1 & \\     
 9230 &  $>$1    & Strong emission \\   
 9724 &  $>$1    & Very weak  \\          
10265 &  $\sim$1 & \\   
10341 &  $\sim$1 & \\   
32469 &   $>$1    & \\   
33918 &  $>$1    & \\   
40983 &  $>$1    & \\   
41969 &  $\sim$1 & \\   
42073 &  $>$1    & \\   
43041 &  $>$1    & \\   
44665 &  $>$1    & \\   
45840 &  $>$1    & \\   
47145 &  $>$1    & \\   
47421 &  $>$1    & \\   
48011 &  $>$1    & \\   
48060 &  $>$1    & \\   
48128 &  $>$1    & \\   
49509 &  $>$1    & \\   
49680 &  $>$1    & Strong emission \\   
49942 &  $>$1    & \\     
50371 &  $>$1    & \\   
50861 &  $>$1    & Strong emission \\   
51515 &  $>$1    & \\   
51871 &  $>$1    & Strong emission \\    
51963 &  $>$1    & Very weak \\   
52006 &  $\sim$1 & Very weak \\     
54284 &  $\sim$1 & Weak \\     
54733 &  $>$1    & Weak \\     
54789 &  $>$1    & \\   
55031 &  $>$1    & Strong emission\\   
\\      
\hline      
\end{tabular}      
\end{center}      
\end{table}

\subsubsection{Shifts and line asymmetry}      
   
In the UVES spectra we derived the position of the H$\alpha$ absorption line    
by defining the center of the line as  the bisector of the profile at half 
maximum intensity, and the core of the line as the minimum intensity 
interpolated by a local parabolic profile.    
     
We have then measured the center and core shifts with respect to the nearby    
photospheric lines, and assumed that only shifts larger than 3$\sigma$    
(i.e. $\sim$ 2 kms$^{-1}$) may be considered significant.      
Of the 20  stars observed with UVES, 8  show significant coreshifts,    
7 of them being blueshifted (indicative of outward motion in    
the layer of the atmosphere where the H$\alpha$ line is formed) and 1 of them    
being redshifted (indicative of a downward flow). The blue-shifted ones are 
all brighter than $\log L/L_{\odot}\sim$2.88 and represent $\sim$54\% of 
the UVES stars brighter than this value; the star with a red-shifted 
H$\alpha$ core is about 3 mag fainter.  
Only two stars show  possibly significant centre shifts, both redward: 
one is the same star that shows also a redshifted core, the other one 
shows no significant coreshift. 
There is no apparent correlation between the occurrence of a blue coreshift    
and the asymmetry of the H$\alpha$ emission wings: of the 7 stars    
where a blue coreshift was detected,  2 have B/R$>$1, 3 have  B/R$<$1, 1 has     
B/R$\sim$1 and 1 does not show emission wings. Five other stars exhibiting    
blue asymmetry do not show any significant coreshift.    
The GIRAFFE spectra have a resolution about a factor 2 worse than the UVES      
data, and the precision is hardly good enough to reveal coreshifts such 
as those detected in the UVES spectra, hence we did not try to estimate them.     

Lyons et al. (1996), thanks to the higher 
resolution (R=60,000) of their spectra, were able to detect and measure 
significant  H$\alpha$ coreshifts in $\sim$50\% of stars brighter than 
$\log L/L_{\odot}\sim$2.5. We cannot reach this limit with our UVES data, 
and GIRAFFE's resolution is not adequate for reliable measures of such 
small shifts. Therefore it appears that $\log L/L_{\odot}\sim$2.5 is the 
threshold for the occurrence of both H$\alpha$ emission (our present results) 
and H$\alpha$ coreshifts (Lyons et al. 1996), at least based on the presently 
available data.  

In all cases the shifts are $\le$10 kms$^{-1}$, as it was found also by 
Smith \& Dupree (1988) from a sample of 52 metal-poor field red giants 
and using echelle spectra of similar resolution.

\subsection{Na {\sc i} D lines}

\begin{figure}      
\begin{center}     
\includegraphics[width=9cm]{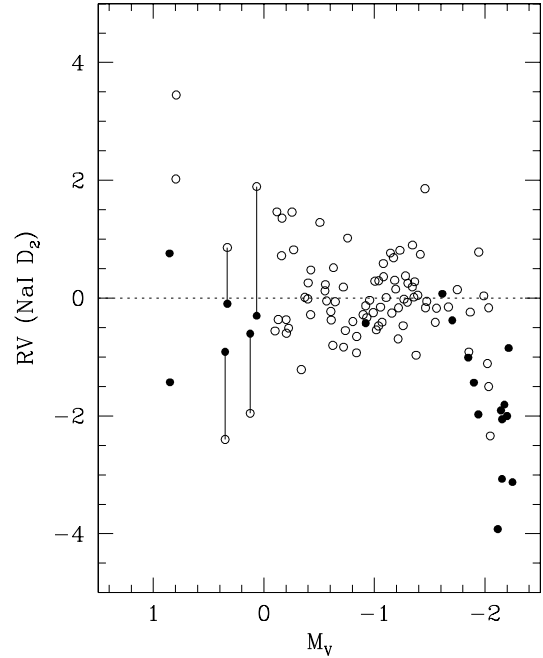}       
\caption{The coreshifts of the Na {\sc i} D$_2$ lines with respect to the    
photospheric     
LSR, as a function of the V magnitude. The filled circles indicate the      
stars observed with UVES, the open circles indicate the stars observed with   
GIRAFFE. Four stars have been observed by both, the two measures are 
connected by a line. Typical errors of the Na {\sc i} D coreshifts are    
$\pm$ 0.6 kms$^{-1}$ for the UVES data and  1.5 kms$^{-1}$ for the GIRAFFE    
data, except for the faintest stars where the errors are larger.        
 }      
\label{nad_shift}      
\end{center}     
\end{figure}      
     
In addition to the 20 stars that were observed with UVES,     
82 more stars  were observed with GIRAFFE/MEDUSA and setup HR12.       
The contributions from the interstellar Na lines are clearly separated,    
since the cluster heliocentric mean radial velocity is $\sim$ 100 kms$^{-1}$.      
     
We have measured the coreshifts of both Na {\sc i} D$_1$ and D$_2$ lines    
with respect to the photospheric LSR, and in Fig. \ref{nad_shift} we show the 
shifts    
of the D$_2$ line (which are generally larger, i.e. more negative,     
than those of the D$_1$ line) as a function of magnitude for all our stars.    
As already commented by Lyons et al. (1996), ``this suggests that, in most    
cases, there is an outward-increasing velocity gradient in the atmosphere    
of these cluster giants (although often for individual stars there is no    
significant velocity difference between the regions of the atmosphere where    
the Na D$_2$ and D$_1$ line cores are formed).''    
Typical errors    
of these measures are $\pm$ 0.6 kms$^{-1}$ for the UVES data and  $\pm$    
1.5 kms$^{-1}$ for the GIRAFFE data, with slightly worse values for the    
faintest stars of our sample. We note that,  within these errors, we do    
not detect any significant (i.e. $\ge 3\sigma$) coreshift in stars fainter 
than $M_V \sim -1.8$ (i.e. $\log L/L_{\odot}\ge$2.9).   
Considering  only the UVES data, 8 out of the 11 stars brighter than this 
value  show a significant negative coreshift, in very good agreement    
with the occurrence of the H$\alpha$ emission wings. 
    
So,  the luminosity limit of $\log L/L_{\odot}\sim$2.9 for the onset of 
significant  Na D$_2$ coreshifts is the same as the value found by Lyons 
et al. (1996), but our detection frequency ($\sim$73\%) is somewhat higher 
than theirs ($\sim$50\%).     
Also, our values for the coreshifts of the D$_2$ line reach at most    
--4 kms$^{-1}$, whereas Lyons et al. (1996) found up to $\sim$ --8 kms$^{-1}$    
in a few stars of the globular cluster M13 and one star in M55.     
This may be a real effect, or might be at least partly due to the resolution 
of our spectra that is a factor $\sim$1.3 lower than  Lyons et al.'s.   
     
Equivalent widths of the Na {\sc i} D lines have been measured for all    
stars, and the  D$_2$ line is the stronger of the Na {\sc i} D pair,    
hence is formed higher in the  atmosphere than the D$_1$ line (as also    
suggested by the larger velocities of the coreshifts).     
The values of the individual equivalent widths are not presented here, as 
they are used elsewhere (Carretta et al. 2003a,b) to perform a detailed 
Na abundance analysis. 
 
Lyons et al. (1996) discuss in some detail the relationship between the 
coreshift and equivalent width of the Na D$_2$ line for their total sample
of 63 stars in 5 globular clusters, and note that significant coreshifts are
found only for stars with EW(D$_2$)$\ge$350 m\AA, with a frequency of 
about 45\%. 
In all of their clusters the variation of the EW(D$_2$) is rather large, 
except in one cluster (M55) where all values of the EW(D$_2$) clump 
below 350 m\AA, 
and no significant coreshifts are detected. This strength threshold, however, 
does not seem to represent a physical threshold for the onset of mass-motion 
phenomena, since in the same M55 stars there is evidence for mass motions 
via H$\alpha$ coreshifts and/or asymmetric H$\alpha$ emission. 
  
It is worth mentioning here that the strength of the D$_2$ line 
can vary significantly from star to star at any luminosity level due to 
intrinsic variations of the Na abundance, as it has been found in our 
NGC 2808 stars, as well as in other RGB stars previously studied in the 
globular clusters M13, M5, M15 and M92 (cf. Carretta et al. 2003a,b for a
discussion). 
Based on Lyons et al. (1996) results, these variations in Na abundance 
within the same cluster RGB stars might lead to miss a significant fraction 
of D$_2$ coreshifts if the corresponding  equivalent width happened to 
fall below the detection threshold of about 350 m\AA. 
Therefore, the use of the Na D$_2$ line negative coreshifts as indicators 
of mass motions in the atmospheres of red giants, as good as it may be,  
could underestimate the real frequency of this phenomenon because of this 
effect. 
  
If chromospheric activity were at work, one might expect that the entire 
atmospheric structure is somewhat affected. 
The onset of H$\alpha$ emission seems to occur at $\log L/L_{\odot}\sim$2.5
and $T_{\rm eff}\sim$4400 K. It is interesting to note that, starting
approximately at this value of temperature, the Na abundances derived
by Carretta et al. (2003a,b) for hotter stars tend to be systematically
lower,  possibly suggesting some degree of line filling.
Also the Fe abundances are slightly smaller, as if the region where
these line form were hotter than predicted by the models.

\subsection{Ca {\sc ii}  K lines}      
     
The Ca {\sc ii} H and K lines are formed in the chromospheres of cool stars    
and appear as a deep absorption doublet with a central emission core that 
may be itself centrally reversed.  These lines are  often used to identify 
mass motions and the presence of circumstellar material in luminous stars 
via their asymmetries (see e.g. Reimers 1977a,b; Dupree 1986), but of course 
the presence of the chromosphere complicates their interpretation, as 
well as for the Na and H$\alpha$ lines, because of the similarity 
between the spectral signatures of chromospheres and mass loss.  

We have observed 83 stars with the GIRAFFE HR02 setup centered on 
the Ca {\sc ii} H and K lines: this is the first time that such a large
collection of  Ca {\sc ii} H and K data for stars belonging to a globular
cluster is available. Previous investigations of chromospheric activity 
in globular cluster giants using Ca {\sc ii} K data were limited  
to two stars in NGC 6752 (Dupree et al. 1994). Our data show a well  
defined Ca {\sc ii} reversal in several stars, and confirm the widespread 
presence of chromospheres among RGB stars of globular clusters. 
       
\begin{figure}     
\begin{center}      
\includegraphics[width=9cm]{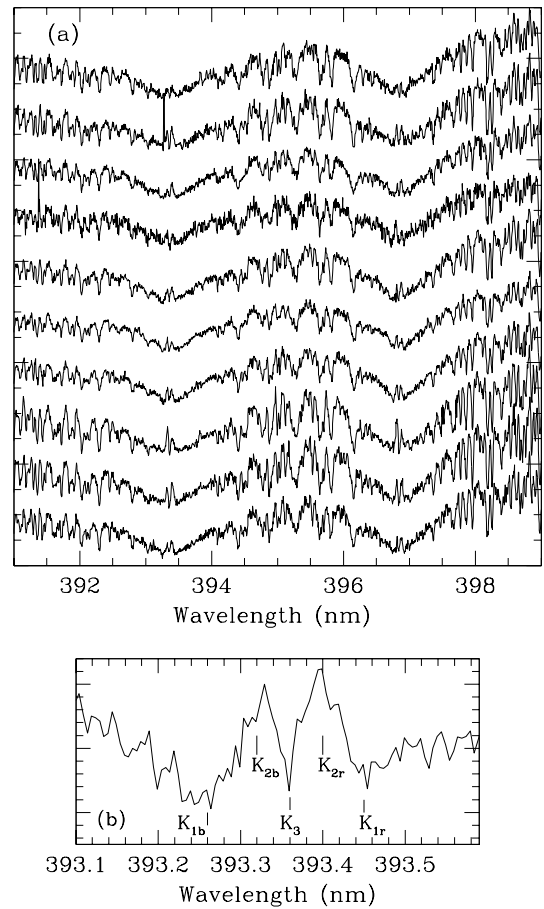}       
\caption{Parts of spectra containing the Ca {\sc ii} H and K lines,     
for some of the brightest stars we have observed.      
The insert shows the zoomed part of the K line for  star \#51499.       
 }      
\label{caii}      
\end{center}     
\end{figure}

We display in Fig.  \ref{caii} the normalised region of the spectra 
containing the     
Ca {\sc ii} H and K lines for some of the brightest stars we have observed.     
The insert shows the bottom of the K line for star \#51499, zoomed to      
show the $K_1$,  $K_2$ and $K_3$     
components. If differential motions are present in the line-forming      
region, these peaks may be unequal in intensity and the central line core     
($K_3$) shifted and asymmetric as the line opacity is moved to the blue      
(expansion) or to the red (contraction).      
Expansion causes the red emission peak ($K_{2r}$) to be strenghtened relative      
to the blue emission peak ($K_{2b}$), as clearly seen in Fig. \ref{k3shift}.         
     
\subsubsection{Emission frequency and asymmetries}     
     
We have observed 83 stars with the GIRAFFE/MEDUSA HR02 setup; of these,      
22  show the central emission and reversal features described above.      
They are all brighter than V=14.4 ($\log L/L_{\odot}\sim$2.6), and    
represent approximately 50\% of our observed sample in this upper    
luminosity interval.      
     
For these stars we have measured the relative strength of the two peaks      
($K_{2b}$ and $K_{2r}$) in the profile of the $K_{2}$ emission reversal,      
usually denoted by B/R (i.e. the ratio of the intensity of the      
short-wavelength to the long-wavelength peaks). The B and R intensities were estimated by fitting a      
Gaussian profile to both   $K_{2b}$ and $K_{2r}$ components of the emission     
(see e.g. Dupree \& Smith 1995). Only when the S/N was too poor, the relative      
intensities were estimated by eye.      
When outward motions are present in the line-forming region, the      
intensity ratio B/R$<$1 due to the increased opacity on the short-wavelength     
side of the line. In our sample, the onset of the B/R$<$1 (red) asymmetry    
seems to occur at $\log L/L_{\odot}\sim$2.87, and applies to about 
75\% of the stars brighter than this value.       
     
We list in Table 5 these stars and corresponding B/R emission asymmetry.      
      
\begin{table}      
\begin{center}      
\label{t:abu1}      
\caption[]{Parameters of the Ca {\sc ii} K line. The $K_3$ shifts are  
measured relative to the nearby photospheric absorption lines.       
}       
\begin{tabular}{rccr}      
\hline\hline      
\\      
Star ID. & B/R   & logW     & $K_3$ shift   \\    
         &       &kms$^{-1}$& kms$^{-1}$   \\   
\hline      
\\   
 9992 & $<$1 & 1.89 & --6.71  \\   
10681 & $<$1 & 2.00 & --4.88  \\   
30927 & $>$1: & 1.98 & --1.45  \\   
37872 & $<$1 & 1.86 & --5.03  \\   
43561 & $>$1 & 1.97 & --5.49  \\     
45162 & $<$1: & 1.99 & --1.30  \\   
46099 & $<$1 & 1.93 & --8.47  \\   
46580 & $<$1 & 1.89 & --5.03  \\   
46726 & $>$1 & 1.83 & +3.36   \\   
47031 & $>$1 & 1.80 & --3.97  \\   
47606 & $>$1: & 1.93 & --6.25  \\   
48889 & $<$1 & 1.91 & --5.19  \\   
50119 & $<$1 & 1.87 & --0.38  \\   
50681 & $<$1 & 1.95 & --6.10  \\   
50761 & $>$1 & 2.06 & --2.29  \\   
51454 & $<$1 & 1.90 & --7.70  \\   
51499 & $<$1 & 1.99 & --6.25  \\   
51930 & $>$1 & 1.90 & +12.89  \\   
52048 & $>$1 & 1.87 & +0.69  \\   
53284 & $<$1 & 2.03 & --12.20  \\   
56536 & $<$1 & 1.90 & +0.38  \\   
56924 & $>$1: & 1.83: & +6.41  \\   
\\      
\hline      
\end{tabular}      
\end{center}       
\end{table}

\subsubsection{Velocity shifts of the $K_3$ reversal}     
     
The $K_3$ self-reversal in the center of the $K_2$ emission line gives      
a direct measure of the velocity in the chromosphere at the highest point      
of Ca line formation. Velocities of the $K_3$ reversal have been measured      
relative to the nearby photospheric absorption lines.         
Typical uncertainties of these measures amount to $\pm$ 1.5 kms$^{-1}$,
and we may assume again that only shifts larger than 3$\sigma$    
(i.e. $\sim$ 4.5 km$s^{-1}$) can be considered significant.          
The values of the $K_3$ velocity shifts are listed in Table 5.       
They are more frequently negative than positive, mostly negative for the      
most luminous stars, and less than 15 kms$^{-1}$ for all stars listed in    
the table. This value is much less than is needed for escape from the 
photosphere, but mass loss cannot be excluded if a sufficiently high 
acceleration were attained at large distances from the star, depending 
on the acceleration mechanism.        
     
\begin{figure}     
\begin{center}      
\includegraphics[width=9cm]{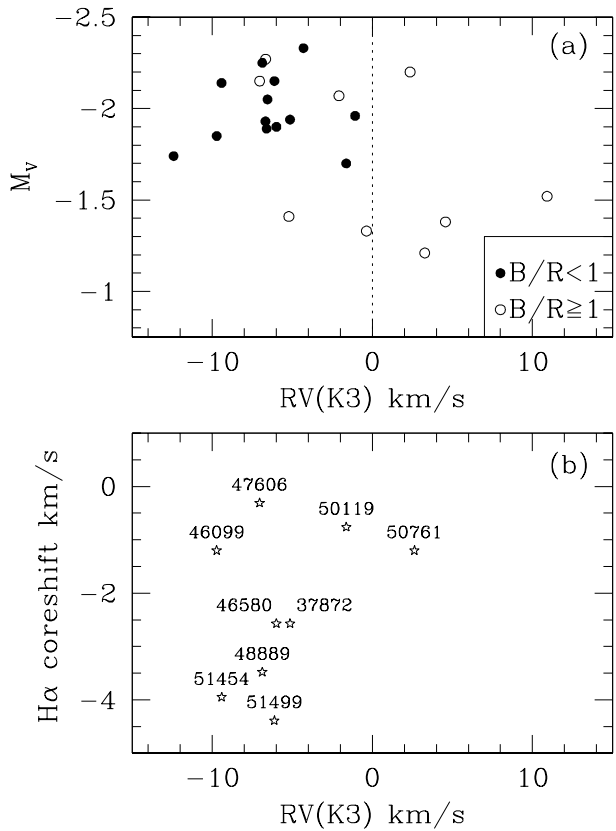}       
\caption{Panel a): $K_3$ velocity shifts as a function of luminosity. 
Stars with red asymmetry in the $K_2$ components  (i.e. B/R$<$1) are shown  
as filled circles, stars with blue asymmetry (i.e. B/R$>$1) are indicated 
as open circles. Panel b): H$\alpha$ coreshift vs. $K_3$ velocity shifts 
for the 9 stars that have both sets of measures. 
 }      
\label{k3shift}      
\end{center}     
\end{figure}

Negative $K_3$ velocity shifts are taken to indicate that there is an      
outflow of material in the region of formation of the $K_3$ core reversal.      
We show in Fig. \ref{k3shift}a a plot of the $K_3$ velocity shifts as a function of      
$M_V$ magnitude, where stars with red and blue $K_2$ emission asymmetries are      
indicated with different symbols: as expected, negative $K_3$ velocity shifts     
are more often associated with red asymmetries, reinforcing the indication    
of  outward motions. These features are also found, to a much larger extent, 
in Population I bright giants where they suggest the presence of strong 
cool winds (cf. Dupree 1986). In Population II giants these winds, if present, 
would appear to be much weaker based on these diagnostics.    
In Fig. \ref{k3shift}b we show the behaviour of the H$\alpha$ 
coreshifts relative to the $K_3$ velocity shifts for the 9 stars that 
have both sets of measures: 7 out of 9 stars define a clear tendency for 
more negative H$\alpha$ coreshifts to be associated with more negative 
$K_3$ coreshifts, stressing again the indication of outward motion; 
two stars (46099 and 47606) fall out of this trend, and this could be 
due to observational errors or variability in the H$\alpha$ profile, 
without pretending however to overinterpret the data. 

Among our sample stars, the onset of negative $K_3$ coreshifts occurs 
at $\log L/L_{\odot}\sim$2.8, i. e. at a slightly lower    
luminosity than the onset of red asymmetry, and it applies to about 89\%    
of the stars brighter than this value. This threshold is $\sim$0.2 mag  
fainter than the threshold estimated by  
Dupree \& Smith (1995) and Smith \& Dupree (1992) from metal-poor field  
red giants, who find significant negative $K_3$ shifts only among stars  
more luminous than $M_V$=--1.7 (i.e. $\log L/L_{\odot}\sim$2.88), in a  
proportion of $\sim$73\%.

\subsubsection{The Wilson-Bappu effect}     
     
The full width W of the emission reversal has been shown to be related    
to  the absolute magnitude $M_V$ of a red giant stars (Wilson \& Bappu 1957).      
The most recent calibration based on a sample of 119 nearby stars 
($M_V$ derived from the {\it Hipparcos} database and W($K_2$) from  high 
resolution spectra)  has been provided by Pace et al. (2003), where an 
extensive discussion and references to the many previous studies are also 
given. 
The most recent analogous survey of metal-poor giants was presented 
by Dupree \& Smith (1995), who analysed 24      
metal-poor field red giants and found that, on average, they are more 
luminous than Population I giants at a given value of the Wilson-Bappu 
width W.      
     
\begin{figure}     
\begin{center}      
\includegraphics[width=9cm]{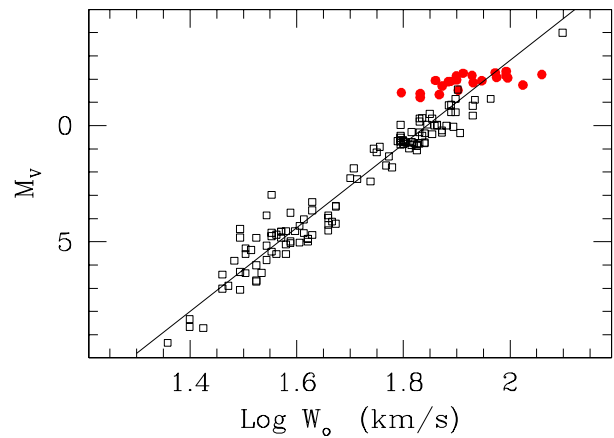}       
\caption{Wilson-Bappu effect: $\log$W as a function of $M_V$. The open squares 
show the Population I giants, and the filled circles show the NGC 2808 giants 
listed in Table 5. The line is the  relation for Pop. I giants derived by      
Pace et al. (2003), $M_V=33.2-18\log W$ (r.m.s. error of the fitting is 
0.6 mag).     
 }      
\label{wb}      
\end{center}     
\end{figure}

We have measured W for 22 of our stars following the same criteria      
described by  Dupree \& Smith (1995) and Pace et al. (2003), and we list      
the values of $\log$W in Table 4. We then show in Fig. \ref{wb} the values of 
$\log$W as a function of $M_V$,  and compare them with the  calibration 
for Population I stars given by Pace et al. (2003).  
In very close agreement with  Dupree \& Smith 
(1995) we also find that the luminosity distribution of our Population II 
giants is nearly flat and mostly contained within the interval 
$< M_V > \sim -1.9 \pm 0.3$ mag, falling above the bright end of the  
$M_V$-$\log$W distribution for Population I giants.  

In more detail, if we consider only the 17 stars brighter than V=14, they  
have $<V>$=13.55 and $\log <W>$=1.948:  if we enter these values in the Pop.I 
relation we find  $<M_V>$=--1.86 and (m--M)$_V$=15.41. These estimates compare 
very well with  $<M_V>$=--2.04 based on the assumed distance modulus 
(m--M)$_V$=15.59, given the accuracy of this method on absolute magnitude 
(hence distance) determinations. 
This implies that, at least for these bright giants up to the metallicity of 
NGC 2808 ([Fe/H] $\sim$ --1.2), the dependence of the Wilson-Bappu law on 
metallicity seems smaller than estimated by Dupree \& Smith (1995). 
Their statement   ``Use of the observed Ca {\sc ii} K line width to derive 
an absolute magnitude with a calibration based on Population I stars will 
underestimate the true $<M_V>$ of a metal-deficient giant'' is still true, 
but the error does not seem dramatic.  
   
\section{Summary and Conclusions}

We have observed a total of 137 RGB stars in the globular cluster NGC 2808 with      
the multi-fibre spectrograph FLAMES, and have measured  the Ca {\sc ii} K,     
Na {\sc i} D and H$\alpha$ lines for 83, 98 and 98 stars respectively. 
We have searched for evidence of mass motions in the  atmospheres using 
diagnostics such as absorption line coreshifts and asymmetries in emission 
components. 
This is the first time that such a large sample of 
globular cluster RGB stars have been observed in all the major optical
lines that are normally used to study the presence of chromospheres 
and/or mass motions in the atmospheres. 
 
Four of these stars are not cluster members, as their radial      
velocities  differ by more than 3$\sigma$ from the cluster average velocity; 
for another star the metal abundance is abnormally large, suggesting either 
that the star is non-member or that it is an AGB star (hence the physical 
parameters used to calculate the metal abundance are incorrect).     
     
From the remaining sample, we have obtained the following results:  
\begin{itemize} 
 
\item 
After subtracting a supposedly pure H$\alpha$ absorption template profile,   
we detect evidence of H$\alpha$ emission down to $\log L/L_{\odot}\sim$2.5
for $\sim$72\% of stars; this proportion increases with luminosity and 
becomes $\sim$94\% ($\sim$95\%) among stars brighter than 
$\log L/L_{\odot}=$2.7 (2.9). Compared to all previous studies, including 
those based on higher spectral resolution such as Lyons et al. (1996), 
we have set the detection threshold for H$\alpha$ emission to a fainter 
limit and increased the fraction of stars where emission could be detected. \\
The nature of this emission is not assessed definitely yet:  our data 
favour a chromospheric  rather than a circumstellar origin, however H$\alpha$ 
emission may be present in either stationary or moving chromospheres. 
Line coreshifts and asymmetries should in principle help distinguish 
between these two possibilities. For the coreshifts we have used only 
the 20 stars observed with UVES because of the better spectral resolution. 
We find that 7 out of these 20 stars, all brighter than 
$\log L/L_{\odot}\sim$2.88, show significant H$\alpha$ coreshifts. 
In all cases the shifts are less than 9 kms$^{-1}$, and negative   
(i.e. indicative of outward motion in the 
layer of the atmosphere where the  H$\alpha$ line is formed). \\ 
Asymmetry of the H$\alpha$ emission wings, indicated as the ratio B/R of 
the blue and red wing intensities, is mostly blue and does not seem to be 
correlated with the absorption coreshifts.  However, its  interpretation is 
likely complex as it is well known from previous studies that both the 
intensity of the emission wings and the sense of the asymmetry can be 
variable with time.  
 
\item 
Na D$_1$ and D$_2$ lines were observed for 20 stars with UVES and 82 stars 
with GIRAFFE. D$_2$ line coreshifts were measured for all of them, but 
significant negative values (i.e. $\ge 3\sigma$) were found only among 
UVES stars 
brighter than $\log L/L_{\odot}\sim$2.9, with a detection frequency of 
about 73\%.  This luminosity threshold for the onset of significant 
Na D$_2$ negative coreshifts is the same as the value found by Lyons et al. 
(1996), whose detection frequency is however of $\sim$50\%.   
The values we find for the D$_2$ coreshifts reach at most --4 kms$^{-1}$, 
whereas Lyons et al. (1996) find values up to --8 kms$^{-1}$; this might 
be a real effect, or might be at least partly due to the lower resolution 
of our data. 
     
Equivalent widths of the Na {\sc i} D lines have been derived for all    
stars, and used in separate papers (Carretta et al. 2003a,b) 
to perform a detailed  abundance analysis. 
Because of possible intrinsic variations of Na abundance among RGB stars 
in a given cluster at any luminosity level, a significant  fraction of 
negative  D$_2$ line coreshifts might escape detection if the corresponding 
EW(D$_2$) is smaller than the detection threshold ($\sim$ 350 m\AA), according 
to Lyons et al. (1996) results.

\item 
We have observed with GIRAFFE the Ca {\sc ii} K line for 83 stars, 22 of 
which show the central emission K$_2$ and reversal K$_3$ features. 
The detection threshold for these features is $\log L/L_{\odot}\sim$2.6, 
and involves about 50\% of our observed sample in this upper luminosity
interval. \\
Asymmetry B/R (i.e. the intensity ratio of the  K$_{2b}$ and  K$_{2r}$ 
components) could be detected in about 75\% of stars brighter than  
$\log L/L_{\odot}\sim$2.9, and is mostly red (B/R$<$1) indicating 
outward motion. \\
Velocity shifts of the K$_3$ reversal relative to the photospheric LSR 
have been measured, and are mostly negative indicating that there is an 
outflow of material in the region of formation of the K$_3$ core reversal. 
The onset of negative K$_3$ coreshifts occurs at $\log L/L_{\odot}\sim$2.8, 
i.e. at a slightly lower luminosity level than the  onset of red asymmetry, 
and it applies to nearly 90\% of the stars brighter than this value. 
Compared to previous results on metal-poor field red giants by Dupree \&
Smith (1995) and Smith \& Dupree (1992), we have moved the detection 
threshold of negative  K$_3$ coreshifts about 0.2 mag fainter, and we have 
increased by more than 20\% the fraction of stars that display such 
property. 

\item
We have measured the full width W of the Ca {\sc ii} $K_2$ emission reversal 
in the same 22 stars discussed above, and compared them with the Wilson-Bappu 
relation,  $\log$W as a function of $M_V$, defined by Pace et al. (2003) for 
Population I giants. We find that our Population II giants have a nearly 
constant luminosity $< M_V > \sim -1.9 \pm 0.3$ mag, slightly brighter than  
the $M_V$-$\log$W distribution for Population I giants, confirming the 
results by Dupree \& Smith (1995).   
Using the 17 brightest stars of our sample, we  find that the 
dependence of the Wilson-Bappu law on metallicity is smaller than previously 
estimated.   
 
\end{itemize} 
 
In conclusion, our survey of RGB stars in NGC 2808 searching for mass motion 
diagnostics in their atmospheres has been able to reach fainter luminosity 
thresholds and monitor in much more detail along the RGB than any previous 
study in a given globular cluster. 
This is due to the FLAMES ability of reaching faint magnitudes with good 
S/N and good spectral resolution for a large number of stars simultaneously. 

Although some of our diagnostics (e.g. H$\alpha$ emission) may not provide  
an unambiguous interpretation, other diagnostics give clear indications 
of the presence of both chromospheres and mass outflows in the atmospheres 
of these stars.
However, we did not attempt to derive any estimate of mass loss rate,  
that depends on rather uncertain parameterizations of this phenomenon.

\begin{acknowledgements}      
     
We are indebted to G. Piotto for the photometry and astrometry of our targets,    
and to T. Kinman for interesting comments about the early spectroscopic 
work on NGC 2808. We thank the referee (A.K. Dupree) for useful comments 
and suggestions.       
We thank the ESO staff  (in particular F. Primas) for carrying out the     
observations and the preliminary data reduction.        
This publication makes use of data products from the Two Micron All Sky Survey,     
which is a joint project of the University of Massachusetts and the Infrared     
Processing and Analysis Center/California Institute of Technology, funded by     
the National Aeronautics and Space Administration and the National Science     
Foundation.      
        
\end{acknowledgements}

\vfill\eject     
     


\end{document}